\documentclass[sigconf,authorversion,nonacm]{acmart}
\AtBeginDocument{%
  \providecommand\BibTeX{{%
    \normalfont B\kern-0.5em{\scshape i\kern-0.25em b}\kern-0.8em\TeX}}}

\usepackage{import}
\usepackage{xifthen}
\usepackage{pdfpages}
\usepackage{transparent}
\usepackage{tikz}
\usetikzlibrary{matrix}
\usetikzlibrary{patterns}
\usepackage{pgfplots}
\usepackage{xspace}
\usepackage{algorithm}
\usepackage{algorithmicx}
\usepackage{algpseudocode}
\usepackage{pgfplotstable}
\usepackage{caption}
\usepackage{subcaption}
\usepackage{amsmath}
\usepackage{amsfonts}
\usepackage{amsthm}
\usepackage{booktabs}

\usepackage{booktabs}
\usepackage{longtable}
\usepackage{multirow}
\usepackage{arydshln}
\usepackage{tabularx}
\usepackage{makecell}

\newcommand{\M}{M}
\newcommand{\N}{N}
\newcommand{\K}{K}
\newcommand{\Amat}{\textbf{A}}
\newcommand{\Bmat}{\textbf{B}}
\newcommand{\Cmat}{\textbf{C}}
\newcommand{\Smat}{\textbf{S}}
\newcommand{\abold}{\textbf{a}}
\newcommand{\bbold}{\textbf{b}}
\newcommand{\sij}{s_{ij}}
\newcommand{\cij}{c_{ij}}

\newcommand{\comp}{\emph{Compute}\xspace}
\newcommand{\precomm}{\emph{PreComm}\xspace}
\newcommand{\postcomm}{\emph{PostComm}\xspace}
\newcommand{\init}{\emph{init}\xspace}
\newcommand{\zurich}{Z{\"u}rich\xspace}
\newcommand{\fwname}{SpComm3D\xspace}
\newcommand{\densename}{Dense3D\xspace}
\newcommand{\hnhname}{HnH\xspace}
\newcommand{\dist}{\emph{Dist3D}\xspace}
\newcommand{\disttd}{\emph{Dist2D}\xspace}

\newcommand{\msgab}{m_{\alpha \rightarrow \beta}}
\newcommand{\gmap}{globalMap\xspace}
\newcommand{\lmap}{localMap\xspace}

\begin{document}

\title{\fwname: A Framework for Enabling Sparse Communication in 3D Sparse Kernels}

\author{Nabil Abubaker}
\email{nabubaker@inf.ethz.ch}
\orcid{0000-0002-5060-3059}
\affiliation{%
  \institution{ETH \zurich}
  %\streetaddress{P.O. Box 1212}
  \city{\zurich}
  \country{Switzerland}
}

\author{Torsten Hoefler}
\email{htor@inf.ethz.ch}
\orcid{0000-0002-1333-9797}
\affiliation{%
  \institution{ETH \zurich}
  %\streetaddress{P.O. Box 1212}
  \city{\zurich}
  \country{Switzerland}
}

\begin{abstract}

%Existing 3D algorithms for distributed-memory sparse kernels rely on bulk dense communication that does not utilize the sparsity pattern of the input matrix.
%do not utilize the sparsity pattern of the input matrix for reducing communication.
%Current 3D algorithms for distributed-memory sparse kernels primarily rely on bulk dense communication, failing to leverage the sparsity pattern inherent in the input matrix.

%Existing 3D algorithms for distributed-memory sparse kernels suffer from high communication volumes and high memory footprints due to primarily relying on bulk dense communication.

Existing 3D algorithms for distributed-memory sparse kernels suffer from limited scalability due to reliance on bulk sparsity-agnostic communication. While easier to use, sparsity-agnostic communication leads to unnecessary bandwidth and memory consumption. We present \fwname, a framework for enabling sparsity-aware communication and minimal memory footprint such that no unnecessary data is communicated or stored in memory. \fwname performs sparse communication efficiently with minimal or no communication buffers to further reduce memory consumption. \fwname detaches the local computation at each processor from the communication, allowing flexibility in choosing the best accelerated version for computation. We build 3D algorithms with \fwname for the two important sparse ML kernels: sampled dense-dense matrix multiplication (SDDMM) and Sparse matrix-matrix multiplication (SpMM). Experimental evaluations on up to 1800 processors demonstrate that \fwname has superior scalability and outperforms state-of-the-art sparsity-agnostic methods with up to 20x improvement in terms of communication, memory, and runtime of SDDMM and SpMM.
The code is available at: \url{https://github.com/nfabubaker/SpComm3D}
\end{abstract}

\renewcommand{\shortauthors}{Abubaker and Hoefler}

\maketitle
\thispagestyle{plain}
\pagestyle{plain}

\section{Introduction}
%Sparse kernels such as SDDMM and SpMM are used prominently in machine learning workloads resembled in graph/hypergraph neural network training.
%Full-batch training of these networks necessitates performing sparse kernels efficiently in distributed-memoery setting.

% Existing 2D and 3D algorithms for large-scale sparse kernels do not utilize the sparsity pattern of the input matrix for reducing communication.
% We provide a framework 

Large-scale sparse kernels such as SDDMM and SpMM are the core operations in many scientific computing and machine learning applications.
These kernels involve a sparse matrix with large dimensions, usually an incidence matrix for a graph, and two tall-and-skinny dense matrices.
In scientific computing, SpMM is encountered in iterative solvers, when there are multiple right-hand sides, such as
block conjugate gradient~\cite{blockconjugate},
or blocked eigenvalue algorithms, such as block Lanczos~\cite{blockedLanczos} and block Arnoldi~\cite{blockArnoldi}.
In machine learning and data science, both SpMM and SDDMM became popular for their role in methods used to solve low-rank matrix factorization problems used for recommender systems~\cite{koren2009matrix} such as stochastic gradient descent and alternating least squares.
A recent survey by Besta et al.~\cite{besta2024gnn} demonstrated that SpMM and SDDMM are the backbone of all variants of Graph Neural Networks (GNNs), including Convolutional GNNs and Attentional GNNs.
Therefore, improving the parallel scalability of these kernels is pivotal for the feasibility and success of the underlying application.

% In the last few years, large-scale SpMM and SDDMM became even more popular as being the core operations in Graph Neural Network (GNN) training and inference~\cite{besta2024gnn, tripathy2020}.
% Such kernels usually include a sparse matrix with large dimensions, usually an incidence matrix for a graph, and a tall-and-skinny dense matrix that follows one of the dimensions of the sparse matrix.

% \todo{keep, shorten or remove ?}
% Executing such sparse kernels on a single processor is challenging due to the size of the dense tall-and-skinny  dense matrix.
% As an example, consider a sparse matrix of size $100M\times 100M$ and $1B$ nonzeros. 
% Such matrix has a density of $10^{-6}$ and can be stored using at most 24GB of RAM memory using the COO format and double-precision floating point representations.
% On the other hand, two dense tall-and-skinny matrices each of size $100M\times 50$ require 80GB of RAM memory, usually a burden for a single compute node.
% Therefore, distributed-memory execution of such kernels is a necessity for scaling to such large matrix sizes.

% These operations involve the multiplication of sparse and dense
% matrices, which exhibit inherent parallelism. However, achieving
% efficient scalability on distributed-memory systems remains a chal-
% lenge due to the communication overhead involved in transferring
% data between processors

Executing SDDMM and SpMM, as well as similar kernels, in parallel has been heavily studied and improved.
The popular 2D~\cite{cannon1969, summa} and 3D~\cite{agarwal19953dgemm, solomonik2011communication, cosma} parallelizations of Dense Matrix-Matrix Multiply (GEMM) provide superior scalability through reduced communication volume and message counts compared to the 1D parallelization.
These algorithms inspired the design of parallel algorithms for
SpGEMM~\cite{bulucc2012spgemm}, SDDMM~\cite{hnh2022}, and SpMM~\cite{tripathy2020, selvitopi2021} with high success.
% 2D and 3D algorithms have shown state-of-the-art results when computing sparse kernels such as SpMM~\cite{tripathy2020, selvitopi2021}, SDDMM~\cite{hnh2022} and SpGEMM~\cite{}, conforming with the 2D and 3D algorithms designed for .
2D and 3D algorithms provide nice upper-bounds on the communication bandwidth and latency by dividing the matrix in a structured way leading to well-defined dependency relations that could define the required communication.
%3D algorithms follows the same approach and add another layer of structure by trading-off memory for communication in a class of algorithms called \emph{communication-avoiding}.

\begin{figure*}[ht]
    \centering
    \includegraphics[width=0.80\textwidth]{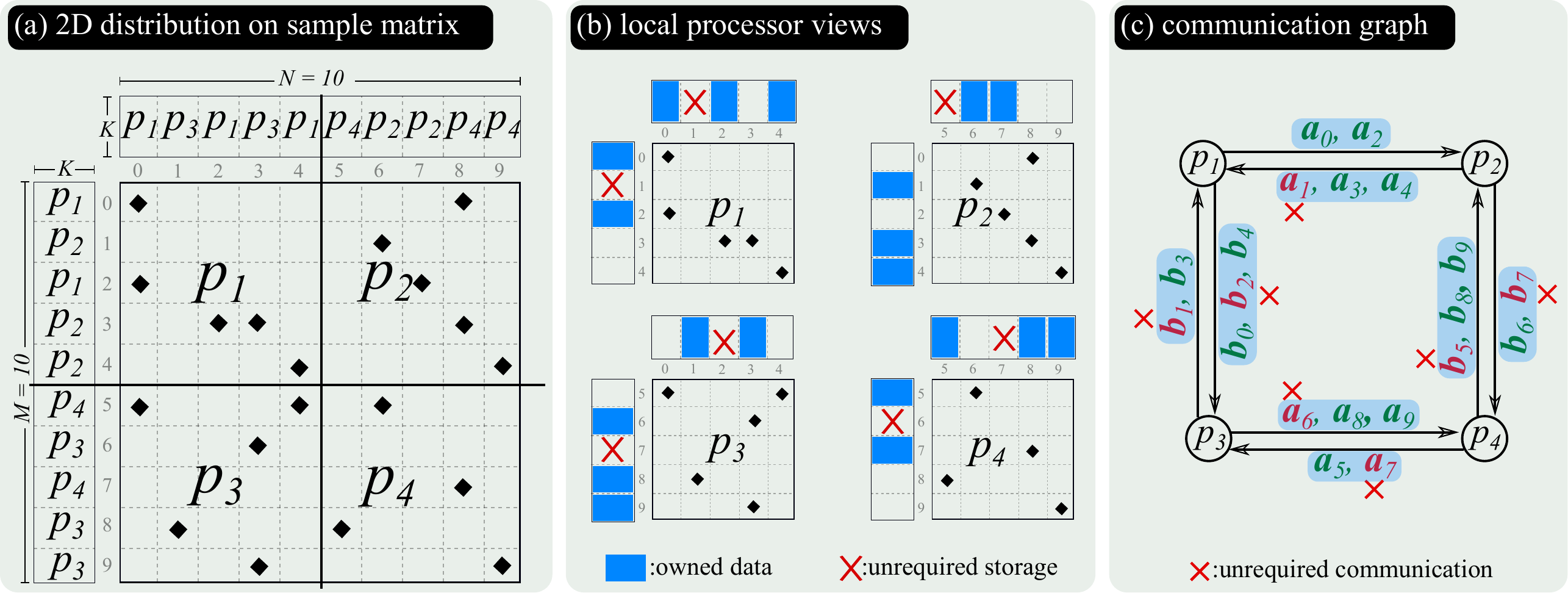}
    \caption{Unnecessary storage and communication in 2D sparsity-agnostic SDDMM. (a) shows a sample 2D SDDMM on a 2$\times$2 processor grid. (b) shows the local view of the sub-problem at each processor; data stored locally but unrequired in computation is marked. (c) shows the communication graph and the unrequired communicated data is marked.}
    \label{fig:localview}
\end{figure*}

Communication in parallel algorithms for sparse kernels is categorized into two categories: sparsity-agnostic bulk communication and sparsity-aware communication.
The former relies on communicating data in bulk, without prior knowledge if the receiving processor does or does not require parts of the data communicated.
This approach's advantages include easier implementation and utilizing existing efficient algorithms for collective operations (e.g., MPI's Broadcast, All-Reduce, and All-Gather). 
On the other hand, it usually involves communicating  unnecessary data that is not used by the receiving processor, especially when the sparsity is very high.
The cost of communicating unnecessary data not only affects the communication bandwidth by adding extra volume, but also requires extra memory for storing this excess data.
We empirically demonstrate that bulk communication becomes too expensive with very large sparse matrices in terms volume and memory overheads.
\figurename~\ref{fig:localview} shows a simple instance of sparsity-agnostic 2D SDDMM and how it stores and communicates data unnecessary for the computation. 

% The sparsity-aware communication offers a bare minimum of communication and memory overheads.
% However, performing sparsity-aware communication can become expensive du usually requires a special attention to implement due 
In this work, we present a novel framework that addresses the scalability challenges of distributed-memory sparse kernels by performing sparsity-aware communication while harnessing the power of 2D and 3D distributions. 
We take advantage of the relatively low, and regular, number of messages in 2D and 3D algorithms to address the latency side, and we perform the bare-minimum required communication to address the bandwidth side of the communication overhead.
The sparsity-aware communication also reduces memory overhead as it enables the storage of only the necessary data required for the computation, thereby improving scalability on large HPC systems.
While there exists 1D~\cite{acer2016spmm, akbudak2018spgemm}, "1.5D"~\cite{mukhopadhyay2023sparse} (communication-avoiding version of 1D), and 2D~\cite{kaya2018partitioning} algorithms in the literature that utilize sparsity-aware communication, to our knowledge, there exists no 3D algorithm that considers sparsity-aware communication.

%In this work, we propose a framework for building 2D and 3D algorithms for sparse kernels with sparsity-aware communication.

% Existing approaches to distributed-memory sparse kernels usually rely on bulk communication, where entire rows or columns of matrices are communicated among processors using collective operations such as broadcast~\cite{selvitopi2021, koanantakool2016, hnh2022, tripathy2020}. 
% While this approach is straightforward to implement, it leads to excessive data transfer, especially for large matrices, thereby limiting scalability.

% In this paper, we present a novel framework that addresses the
% scalability challenges of distributed-memory sparse kernels by leveraging sparse communication. Our framework reduces communication and memory overheads by
% storing and communicating only the necessary data required for the computation, thereby improving scalability on large HPC systems.

Our framework fits any 3D sparse kernel that have a computation phase preceded or followed, or both, by a communication phase.
\fwname does not change the communication based on the sparsity pattern of the input matrix.
In this paper we focus on SDDMM and SpMM and we show how to build the sparsity-aware 3D algorithms for these kernels with \fwname.
%focus on this paper on two important kernels SDDMM and SpMM. For the convenience of the presentation, we mainly consider SDDMM in our explanations and analysis, and later we provide the connection between SDDMM and SpMM.
Our major contributions are summarized as follows:
\begin{itemize}
    \item We design a framework that provide a general environment for building large-scale sparse kernels with 2D and 3D distributions that perform sparse communication and require minimal memory footprint (\S~\ref{sec:spcomm3d}).
    \item Within the framework, we provide several options to perform the sparsity-aware communication revolving around enabling true zero-copy communication in MPI to further reduce memory footprint.
    \item We outline the communication and memory inefficiencies of the existing 2D and 3D algorithms for SDDMM and SpMM, and we carefully define the minimum required communication to be performed for the correctness of the kernel (\S~\ref{sec:SDDMMComm}).
    \item We utilize the new framework for building efficient sparsity-aware 3D SDDMM and SpMM algorithms, and we provide a bottom-up guide to doing so in order to pave the way to build other kernels.
 
\end{itemize}

% \todo{fix this, outdated info+section numbers}
% The rest of the paper is organized as follows: In Section~\ref{sec:relatedwork} we overview the state-of-the-art works and how they differ from our work.
% Section~\ref{sec:bg} covers background on SDDMM and its parallelization methods.
% In Section~\ref{sec:SDDMMComm} we provide thorough analysis of the communication required in the 2.5D SDDMM algorithm and we detail how this communication is satisfied with bulk communication and sparse communication.
% Our framework \fwname is covered in details in Section~\ref{sec:spcomm3d}.
% We provide detailed experimental evaluation in Section~\ref{sec:exp} and we conclude the paper in Section~\ref{sec:conclusion}.

\section{Related Work}
\label{sec:relatedwork}

% \todo{taxonomy figure?}
% \begin{figure}[t!]
% \centering
% \resizebox{\columnwidth}{!}{
% \begin{tikzpicture}[
%     level 1/.style={sibling distance=4cm},
%     level 2/.style={sibling distance=2cm},
%     level 3/.style={sibling distance=1cm}
% ]
%     \node {Communication Type}
%         child {node {Sparsity-Agnostic}
%             child {node {1D}
%                 child {node {SpMM}}
%                 child {node {SDDMM}}
%                 child {node {Random Text 3}}
%             }
%             child {node {2D}
%                 child {node {Random Text 4}}
%                 child {node {Random Text 5}}
%                 child {node {Random Text 6}}
%             }
%             child {node {3D}
%                 child {node {Random Text 7}}
%                 child {node {Random Text 8}}
%                 child {node {Random Text 9}}
%             }
%         }
%         child {node {Sparsity-Aware}
%             child {node {1D}
%                 child {node {SpMM}}
%                 child {node {SpGEMM}}
%                 child {node {SDDMM}}
%             }
%             child {node {2D}
%                 child {node {Random Text 13}}
%                 child {node {Random Text 14}}
%                 child {node {Random Text 15}}
%             }
%             child {node {3D}
%                 child {node {Random Text 16}}
%                 child {node {Random Text 17}}
%                 child {node {Random Text 18}}
%             }
%         };
% \end{tikzpicture}
% }
% \caption{Taxonomy}

% \end{figure}

2D and 3D algorithms impose upper bounds on communication volume in parallel GEMM operation, and on the number of messages in parallel SpGEMM operation, compared to 1D algorithms.
3D algorithms, such as the work of Agarwal et al.~\cite{agarwal19953dgemm}, are considered the communication-avoiding version of 2D algorithms as they create multiple copies of the input data to reduce communication.
``2.5D'' algorithms, first introduced by Solomonik and Demmel~\cite{solomonik2011communication} and later improved by Kwasniewski et al.~\cite{cosma} are similar to 3D algorithms but control the number of copies created according to available memory.
These classes of algorithms are later adapted to SpGEMM:
2D algorithm in the work by Bulu\c{c} and Gilbert~\cite{buluc2008spsumma,bulucc2012spgemm}, 3D algorithm by Ballard et. al~\cite{ballard2013communication}, and 2.5D algorithm by Azad et al.~\cite{azad2016spgemm25d}.
The work by Koanantakool et al.~\cite{koanantakool2016} introduced a new class of algorithms called "1.5D" (i.e., the communication avoiding version of 1D) for the sparse-dense matrix-matrix multiplication (SpDM$^{\textrm{3}}$) operation by making redundant copies of the input matrices on top of 1D distribution.
The communication in all of these algorithms is sparsity-agnostic.
%Parallel algorithms for $N$-order sparse tensors also benefited from 2D/3D to $N$D distributions~\cite{Smith10}.

% 2D and 3D sparse kernels are designed with inspiration from the long-studied 2D and 3D dense matrix-matrix multiplication (GEMM) algorithms.
% 2D algorithms  started with Cannon's algorithm~\cite{cannon1969} and Fox's algorithm~\cite{fox1988}. 
% SUMMA~\cite{summa} is currently considered the state-of-the-art 2D GEMM algorithm.
% 3D algorithms are later introduced firstly by Argwal at al.~\cite{agarwal1995three} and provided lower communication overheads by trading communication for memory, a technique which later became known as \emph{communication avoidance}.
% 2.5D GEMM algorithms, introduced by Solomonik and Demmel~\cite{solomonik2011communication}, replicate the matrices according to available memory.
% COSMA~\cite{cosma} picks the optimal 3D dimensions according to the matrix dimensions to achieve lower communication bounds.

Several works provided 2D and 3D sparsity-agnostic algorithms for SpMM and SDDMM.
Tripathy et al.~\cite{tripathy2020} provided and thoroughly analyzed and compared 1D, 1.5D, 2D, and 3D algorithms for SpMM.
Bharadwaj et al.~\cite{hnh2022} showed how to convert SpMM algorithms to SDDMM, and provided several 2.5D algorithms for SpMM, SDDMM, and FusedMM, a term they coined for the cascade of SDDMM into SpMM, which appears in GNN training and inference.
Kannan et al.~\cite{kannan2017mpi} implemented 2D SpMM within the scope of distributed-memory non-negative matrix factorization algorithm called MPI-FAUN.
%Kaya et al.~\cite{kaya2018partitioning} later improved upon MPI-FAUN by employing sparsity-aware communication for the 2D SpMM and exploring different partitioning strategies.
Selvitopi et al.~\cite{selvitopi2021} provides thorough analysis and efficient RDMA-based implementations of several configurations of 2D SpMM.

\sloppy{Sparsity-aware communication, or communication that follows the sparsity of the input matrix, has been previously implemented and analyzed in the context of several 1D algorithms, prominently present in works that utilize graph/hypergraph partitioning for reducing communication overhead in 
kernels such as sparse matrix-vector multiplication (SpMV)~\cite{ccatalyurek2010two, kaya2014, kayaaslan2018-1.5D, demirci2021partitioning}, SpMM~\cite{acer2016spmm} and SpGEMM~\cite{akbudak2014techreport, akbudak2014simultaneous, akbudak2018spgemm}.
Kaya et al.~\cite{kaya2018partitioning} implemented a 2D sparsity-aware SpMM algorithm for non-negative matrix factorization on distributed-memory systems.
Abubaker et al.~\cite{abubaker2023minimizing} implemented a 1D SDDMM algorithm for asynchronous for asynchronous SGD used in matrix factorization.
In a recent technical report, Mukhopadhyay et al.~\cite{mukhopadhyay2023sparse} utilize sparsity-aware communication for 1.5D SpMM on GPUs. 
To our knowledge, our work is the first to implement sparsity-aware communication in 2.5D/3D algorithms.}

% A recent technical report by Mukhopadhyay~\cite{mukhopadhyay2023sparse} presents an algorithm for sparsity-aware 1.5D SpMM. 
% Their work is similar to our work in regards to utlizing sparsity for communication, however it differs in the following aspects: (i) their work considers 1.5D algorithm, a 1D distribution ($P/c$) replicated $c$ times, for SpMM on GPUs, whereas our work considers both 2D and 3D algorithms, (ii) they consider an $s$-step algorithm, where $s = \frac{P}{Z^2}$ for SpMM, whereas we consider single-step algorithms. (iii) their work targets communication between GPUs via NCCL, whereas our framework deals with general inter-node communication via MPI and does not have any assumption about the device on which the computations will take place. (iv) their work computes the sparsity-aware communication information on-the-fly (i.e., during the execution of the main algorithm), whereas we perform this in a dedicated setup phase prior to starting the main algorithm.

\section{Preliminaries}
\label{sec:bg}

\subsection{Notations}

\fwname builds a logical 2D or 3D Cartesian processor grid and distributes the input sparse matrix/matrices onto this grid in a structured fashion.
The number of processors in denoted by $P$, the set of all processor by $\mathcal{P}$, and an arbitrary processor in $\mathcal{P}$ by $p_{\alpha}$.
For 2D grids, we use the notation $P_{x,y}$ to indicate the processor at location ($x$,$y$) of the grid.
Similarly, for 3D grids, $P_{x,y,z}$ is located at ($x$, $y$, $z$) in the grid.
We use the Matlab notation to address a whole row block, column block, or a slice of the processor grid (e.g., $P_{x,:,z}$ means the $x$th row block $\{P_{x,1,z}, P_{x,2,z}, \ldots, P_{x,Y,z}\}$ of the 3D grid and $P_{:,:,z}$ means the $z$th slice (2D grid) of the 3D grid). 

$\Smat$ and $\Cmat$ always refer to sparse matrices, whereas $\Amat$ and $\Bmat$ always refer to dense tall-and-skinny matrices.
$\Smat(i,:)$ and $\Smat(:,j)$ denotes the $i$th row and $j$th column of $\Smat$, respectively, and a nonzero element by $\sij$. 
The $i$th row in $\Amat$ or $\Bmat$ is denoted by $\abold_i$, $\bbold_i$.
The functions $nnz(\cdot)$, $nrows(\cdot)$ and $ncols(\cdot)$ are respectively used to denote the number of nonzero elements, number of rows, and number of columns of a (sub)matrix.
We use $X$, $Y$ and $Z$ as dimensions for the 2D or 3D processor grid, and $M$, $N$, and $K$ are used as dimensions for the matrices. $K \ll (M,N)$ is always the number of columns for the tall-and-skinny matrices.
% The number of processors in denoted by $P$, the set of all processor by $\mathcal{P}$, and an arbitrary processor in $\mathcal{P}$ by $p_{\alpha}$. 

\subsection{SDDMM and S\lowercase{p}MM}

Sampled Dense-Dense Matrix Multiplication (SDDMM) \mbox{$\Cmat=\Smat \odot \Amat \Bmat^T$}, where $\odot$ means element-wise multiplication, involves four matrices: $\Smat$ and $\Cmat \in \mathbb{R}^{\M\times \N}$  are respectively the sparse sampling and output matrices. $\Amat \in \mathbb{R}^{\M \times \K}$ and $\Bmat \in \mathbb{R}^{\N \times \K}$ are the input dense tall-and-skinny matrices. 
For large $\M$ and $\N$ values, materializing the $\Amat \Bmat^T$ product becomes prohibitive.
Since it will be sampled by the sparse $\Smat$ matrix, most of the elements in this product are not required, and thus can be computer more efficiently element-by-element as follows:
\begin{equation}
	\cij = \sij \times \langle\abold_i, \bbold_j \rangle, \forall (i,j) \in \{(i,j)\mid \sij \ne 0\}
 \label{eq:sddmm}
\end{equation}

% For large matrix dimensions, storing and computing dense ma-
% trices becomes impractical due to memory and computational con-
% straints. Moreover, in SDDMM, the dense $AB$ product is not entirely
% required since it will be sampled by the sparse $S$ matrix, reflecting
% its sparsity pattern. Hence, performing element-wise computation
% can significantly reduce storage and computation costs.

Sparse times Dense Matrix Multiplication (SpMM) $\Amat=\Smat \Bmat^T$ involves $\Smat$, $\Bmat$ as input, and $\Amat$ as output, all of sizes described above. SpMM can be computed row-by-row in serial execution as
\begin{equation}
\abold_i = \sum_{1\le j\le N} \sij \times \bbold_j.
\label{eq:spmm}
\end{equation}
%in order to exploit data locality.

Both SDDMM and SpMM operate on one sparse matrix and two dense tall-and-skinny matrices.
Both operations require taking/outputting two dense vectors per nonzero element. 
In SDDMM, both vectors are taken as input, whereas in SpMM $\bbold_j$ is taken as input and $\abold_i$ is given as output.
Therefore, most of the analyses and algorithms explained afterwords apply to both operations.
Our presentation, figures, and analyses will focus on SDDMM hereafter, and we will explain the difference if required.

To parallelize~\eqref{eq:sddmm} and~\eqref{eq:spmm}, the fine-grain tasks are first identified, then the communication is determined following how these fine-grain tasks are distributed to different processors and the dependency between them.
For SDDMM, we define the fine-grain task as computing one scaled inner product. 
Each fine-grain task requires two vectors (each of size $\K$ words) and one scalar.
For SpMM, we define the fine-grain task as computing one vector scaling as in~\eqref{eq:spmm}.
Each fine-grain task requires one vector of size $\K$ words and a scalar.
%We will later show how these fine-grain task definitions are further divided into sub-tasks when we move from 2D to 3D algorithms.

\subsection{Sparsity-agnostic parallel algorithms for SDDMM and SpMM}
\label{sec:comm}

\begin{figure*}[t!]
    \centering
    \includegraphics[width=0.9\textwidth]{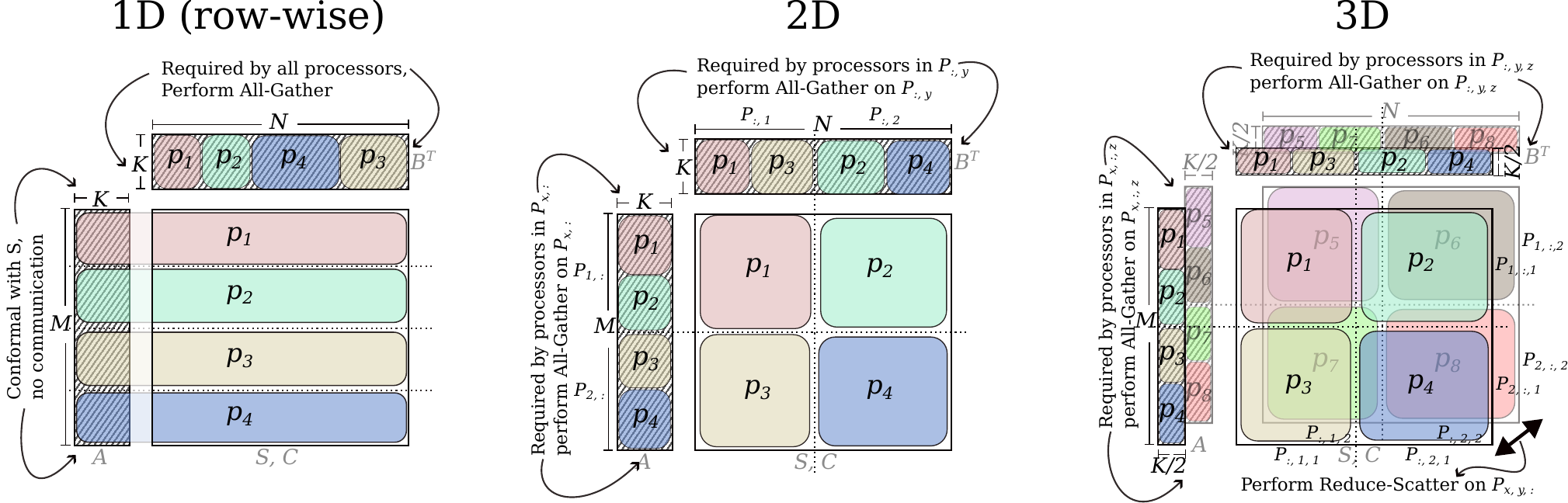}
    \caption{Sparsity-agnostic 1D, 2D, and 3D SDDMM. The communication required stays the same whether $\Smat$ is dense or sparse. 
    %In row-wise 1D, the rows of $A$ are distributed following the rows of $S$, and are required by only once processor, hence not communicated. The rows of $B$, on the other hand, are distributed to different processors and required by every processor thus need to be gathered at every processor before the SDDMM computation takes place.
    }
    \label{fig:topView}
\end{figure*}

Parallel SDDMM and SpMM algorithms are categorized according to how the input sparse matrix is partitioned among processors. 
The partitioning can either be based on granularity or on structure.
Granularity-based categorization can be either fine-grain (nonzero-based) or coarse-grain (row or column-based).
Structure-based categorization relies on the dividing the computational iteration space in a structured manner into 1D, 2D, or 3D shape, which mirrors a partitioning on the dimensions of the matrix itself.
We provide an overview of such algorithms and provide a volume upper bound for the best algorithm in each category.
\figurename~\ref{fig:topView} illustrates 1D, 2D, and 3D algorithms for SDDMM.
Since our concrete goal is to reduce the communication volume via sparsity-aware communication, we analyze the communication of the sparsity-agnostic algorithms in terms of the received volume per processor. 
%Although there are several communication metrics that could be considered, we focus on the received volume per processor.

%Communication avoiding algorithms such as 1.5D and 2.5D for SDDMM and SpMM are inspired by the dense matrix-matrix multiply (GEMM). These algorithms decrease the overall communication bandwidth and latency by trading-off memory for communication.

In 1D algorithms, the sparse matrix $\Smat$ is divided into $P$ parts either row-wise or column-wise, necessitating the communication of one of the dense matrices, not both.
Assuming 1D division on rows, the rows of the dense matrix $\Amat$ are also divided conformably with the rows of $\Smat$. 
Naturally, a processor $p_\alpha$ is assigned the ownership of the dense rows of $\Amat$ that align with its assigned row block of $\Smat$, which necessitates no communication on the rows of $\Amat$. 
On the other hand, the dense matrix $\Bmat$ is likely to be used by all processors thus necessitates communication.
Assuming $\Bmat$ is divided into $P$ parts, and each part is assigned to a single processor, each processor will communicate its part to $P\!-\!1$ processors, leading to approximately \[\Bmat_{size} \times\frac{P-1}{P}\] words of volume to be received per processor.
%See \figurename~\ref{fig:topView} (left).
%This amounts to $O \left( \N\K \times (P-1)\right)$ total volume.
The per-processor memory requirement for storing the dense matrices is \[\frac{\Amat_{size}}{P} + \Bmat_{size}.\]

In 2D algorithms, the sparse matrix is divided into a $\sqrt{P}\times \sqrt{P}$ 2D grid.
Depending on how the dense matrices are divided, different versions of the 2D algorithms emerged~\cite{selvitopi2021}.
We consider the version where $\Amat$ and $\Bmat$ are 1D divided row-wise into P parts, which is well-suited for single-step SpMM and SDDMM algorithms~\cite{kannan2017mpi, selvitopi2021, hnh2022}.

% ~\cite{hnh2022, selvitopi2021, tripathy2020},
% However, unlike 2D algorithms for dense and sparse GEMM, $\Amat$ and $\Bmat$ are divided 1D into $\sqrt{P}$ parts, and each of those parts is further divided into $\sqrt{P}$ parts.
Each of the inner $\sqrt{P}$ parts of $\Amat$ along a row block of processors $P_{x,:}$ is assigned ownership to one of the processors in that block, similarly fo $\Bmat$.
%Some works call this 1.5D distribution~\cite{selvitopi2021}, however we stick to 2D as our naming convention follows the sparse matrix.
The 2D grid partitioning restricts the requirement of a given dense matrix part to only $\sqrt{P}$ processors along the same row or column block.
The communication in 2D algorithms is required along both dimensions.
Each processor receives an approximate of \[ (\Amat_{size} + \Bmat_{size}) \times \frac{\sqrt{P}-1}{P} \] words of volume.
%See \figurename~\ref{fig:topView} (middle).
%The communication in 2D algorithms is required along both dimension, and amounts to $O \left( \N\K \times (\sqrt{P}-1)\right)$ of total volume.
The per-processor memory requirement for storing the dense matrices is \[\frac{\Amat_{size} +\Bmat_{size}}{\sqrt{P}}.\]

3D algorithms, sometimes called 2.5D algorithms~\cite{hnh2022, tripathy2020}, are considered the communication-avoiding version of the 2D algorithms.
These algorithms add an additional dimension $Z$ to the 2D algorithms by running $Z$ instances of the sparse kernel concurrently, where each instance is responsible for computing $\K/Z$ columns of the dense matrix.
This naturally means that the each instance is resembled with a replica of the sparse matrix partitioned according to the 2D scheme into $\sqrt{P/Z}\times \sqrt{P/Z}$ grid.
The columns of the dense matrix are partitioned into $Z$ parts, each part is used exclusively by a separate instance.
With this scheme, each processor receives an approximate of 
\[(\Amat_{size} + \Bmat_{size}) \times \frac{\sqrt{\frac{P}{Z}}-1}{P}\] words of volume.
See \figurename~\ref{fig:topView} (right).
The per-processor memory requirement for storing the dense matrices is \[\frac{\Amat_{size} +\Bmat_{size}}{Z\sqrt{\frac{P}{Z}}}.\]

%With this scheme, each processor receives $2\times \frac{NK}{PZ} \times (\sqrt{P/Z}-1)$ words of data, which amounts to $O \left( \N\K \times (\sqrt{P/Z}-1)\right)$ of total volume.

%With this scheme, each processor receives $2\times \frac{NK}{PZ} \times (\sqrt{P/Z}-1)$ words of data, which amounts to $O \left( \N\K \times (\sqrt{P/Z}-1)\right)$ of total volume.

\section{Sparsity-aware Communication Analysis of 3D SDDMM}
\label{sec:SDDMMComm}

In a row processor block $P_{x,:,z}$, let $\lambda_i$ be the number of processors at which row $\Smat(i,:)$ has at least one nonzero element, and $\Lambda_i$ be the set of such processors.
Assuming that row $\abold_i$ of dense matrix $\Amat$ is owned by one of the processors in $\Lambda_i$, it is required by $\lambda_i-1$ processors.
A similar discussion holds for a column processor block $P_{:,y,z}$ and the dense matrix $\Bmat$.
Then, the total communication volume exchanged in sparsity-aware SDDMM is equal to \[\sum_{i \in [\![1,nrows(\Smat)]\!]} (\lambda_i\! -\!1) + \sum_{j \in [\![1,ncols(\Smat)]\!]} (\lambda_j \! -\!1)\]
In the dense case, $\lambda_i$ and $\lambda_j$ become equal to sizes of the $X$ and $Y$ dimensions of the 3D grid ($\sqrt{\frac{P}{Z}}$) which leads to correct total volume.

The communication requirement modelled by $\lambda$ depends on three factors: (i) the sparsity pattern of $\Smat$, (ii) the number of processors and (iii) how the matrix nonzero elements are distributed to processors.
While factor (i) is vital, we assume the sparsity pattern of $\Smat$ is irregular (i.e., non-uniformly distributed) in order for (iii) to take effect.
With the sparsity-agnostic communication, a row/column of $\Smat$ is divided into $P$ parts incurs $\K(P-1)$ words of communication volume. 
This means that the communication volume scales with how many parts a row/column partitioned into.
With $\lambda$, on the other hand, the volume depends on how many of the $P$ parts has at least one nonzero element. 
As $P$ grows larger, the probability of a certain part of a row/column of $\Smat$ is empty becomes higher.
Therefore, the communication requirement according to $\lambda$ is loosely related to the growing $P$.

\begin{figure}[h!]
    \centering
    \includegraphics[width=0.5\columnwidth]{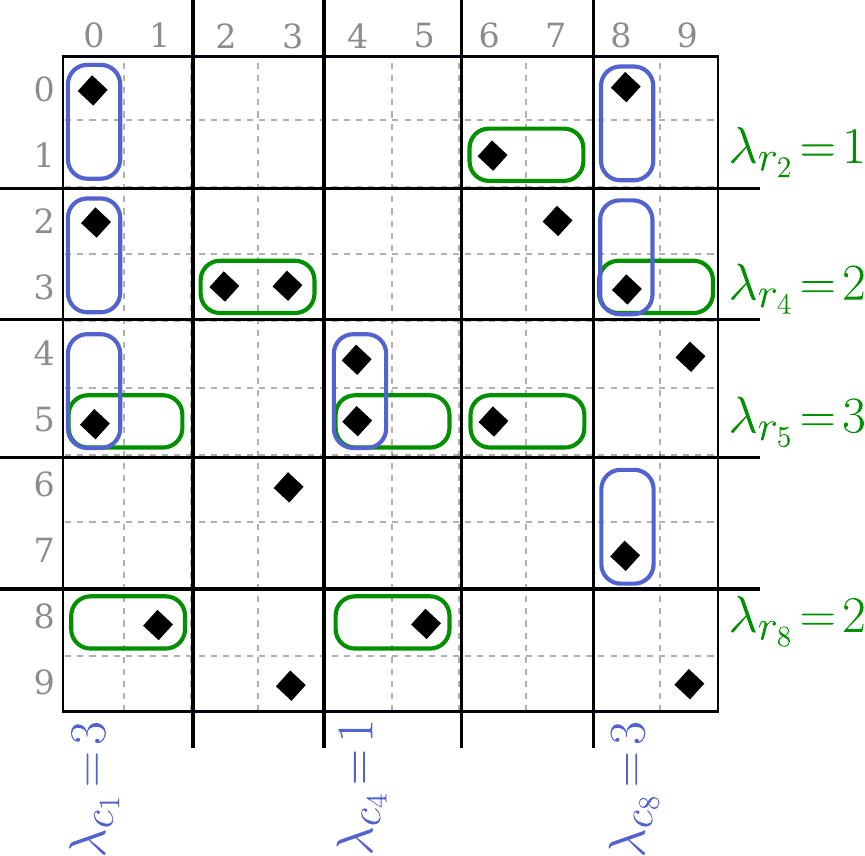}
    \caption{A sample matrix distributed onto 5$\times$5 grid and the $\lambda$ values of some rows/columns.}
    \label{fig:SpM5x5}
    \vspace{-2ex}
\end{figure}

In order to define the received volume per processor, we define two sets $\mathcal{I}_\alpha = \{i \mid \Smat_{\alpha}(i,:) \ne \emptyset \land owner(\abold_i) \ne p_\alpha\}$ and $\mathcal{J}_\alpha = \{j \mid \Smat_{\alpha}(:,j) \ne \emptyset \land owner(\bbold_j) \ne p_\alpha\}$ 
as respectively the sets of row and column indices such that $p_\alpha$ has at least one zero in the row/column of $\Smat$ in those indices and $p_\alpha$ is not the owner of the corresponding dense row.
Then, the volume received by $p_\alpha$ is equal to $\frac{K}{Z}(|\mathcal{I}_\alpha| + |\mathcal{J}_\alpha|)$
% \begin{equation}
%     \frac{K}{Z} \left(\sum_{i \in \mathcal{I}_\alpha} (\lambda_i - 1) + \sum_{j \in \mathcal{J}_\alpha} (\lambda_j - 1) \right)
% \end{equation}
words of volume.  
%While the $\lambda$-based communication requirement is satisfied with sparsity-agnostic bulk communication, our goal is to satisfy it with 

Our goal in this work is to achieve the $\lambda$-based communication requirement defined in this section using efficient communication without extra unnecessary volume. 
Building the structure of this communication for most sparse kernels can be cumbersome, which is why we introduce \fwname  to provide a structured and convenient environment for building sparse kernels with minimum communication.

\section{The \fwname Framework}
\label{sec:spcomm3d}

%\subsection{General structure and assumptions}
The framework has three design goals: (i) communicate only required data, (ii) minimal memory footprint, and (iii) communication-agnostic local computation at each processor to allow utilizing existing efficient algorithms/tools for sparse/dense kernels.

\subsection{General structure and assumptions}
\fwname assumes that the sparse kernel is used in a larger context, and is repeated multiple times. 
The sparsity pattern of the input sparse matrix/matrices remains fixed, while the values might be updated.
\fwname also works under the assumption that there are other dense or sparse data that will be computed or used during the sparse kernel, and this data is updated at each iteration either before or after executing the sparse kernel.

\fwname revolves around a local computation, and communicates/stores the minimum amount of data required for the overall correctness of this computation.
Therefore, computing a sparse kernel with SpComm3D naturally involves three phases: \precomm, \comp, and \postcomm.
\precomm gathers the data required for computation from different processors, while \postcomm communicates partial results to the processors responsible for holding the final value.

\comp is the local computation phase at each processor.
The assumption here is that the local computation is agnostic to the communication and the general structure of the problem, thus should be operating on local data with localized indices.
The \precomm phase ensures that the data used in the \comp phase is correct and up-to-date, bridging the gap between the global and local views of the problem.

Data communicated in \precomm and \postcomm is represented in local memory by a data segment, which may consist of one or more data words and is identified globally using a unique ID typically related to the sparse matrix's global row/column indices.
We call such data segment Data Unit (DU) hereafter.
A DU might be required or updated by several processors but owned by only one, which can be retrieved with $owner(\cdot)$.

As per our assumption that the sparse kernel will be executed multiple times with the same sparse matrix, we minimize the amount of work to be done during the \precomm and \postcomm phases by introducing a setup phase that is executed once.
This phase builds all the communication structures, buffers, and meta information that are used in the communication.
Then, the \precomm and \postcomm phases are used to merely move data between processors. 
A similar philosophy is followed in MPI's persistent communication. 
% Algorithm~\ref{alg:skeleton} shows the general skeleton of setting up and computing a sparse kernel with \fwname.

\subsection{Data distribution}
% \fwname builds a logical 2D ($X\times Y$) or 3D ($X\times Y \times Z$) Cartesian processor grid and distribute the input sparse matrix/matrices onto this grid in a structured fashion.
% %Here, we define how the sparse data is distributed when discussing 2D and 3D algorithms for SDDMM and SpMM.
% %For 2D algorithms, we assume the processors are organized as a 2D Cartesian grid of dimensions $X\times Y$
% For 2D grids, we use the notation $P_{x,y}$ to indicate the processor at location ($x$,$y$) or the grid.
% Similarly, for 3D grids, $P_{x,y,z}$ is located at ($x$, $y$, $z$) in the grid.
% We use the Matlab notation to address a whole row block, column block, or a slice of the processor grid (e.g., $P_{x,:,z}$ means the row block $\{P_{x,1,z}, P_{x,2,z}, \ldots, P_{x,Y,z}\}$ of the 3D grid and $P_{:,:,z}$ means the $z$th slice (2D grid) of the 3D grid). 
% %Hereafter, we use $\mathcal{P}$ to refer to the set of all processors and $P$ to refer to their count.

We assume that the input matrices $\Smat$ is distributed onto the 2D/3D processor grid as follows: the matrix is partitioned into $X\!\times\! Y$ in the row/column dimension spaces.
Each block $\Smat_{x,y}$ of the partitioned matrix is assigned to processor $P_{x,y}$ in a 2D grid, and then partitioned into $Z$ parts in the nonzero space.
That is, $\Smat_{x,y}$ is distributed equally among $Z$ processors in $P_{x,y,:}$ as $\Smat^1_{x,y},\cdots,\Smat^Z_{x,y}$ in a 3D grid.
%Then, the nonzero elements for each ($x$,$y$) block of the $S$ matrix $S_{x,y}$ is distributed to $Z$ parts. 
%This distribution scheme naturally creates a 3D distribution which cab be naturally projected onto the 3D processor grid $P$. 
We refer to this distribution as \disttd or \dist depending on the grid used.

Each matrix block $\Smat^z_{x,y}$ is localized to its processor by localizing row/column indices and removing empty rows/columns.
In order to keep the global information of the local sparse matrix, \fwname stores two arrays along the rows and the columns: $\gmap$ and $\lmap$.
$\gmap$ stores the global index of each local row/column and $\lmap$ stores local indices of the rows/columns in the sub-matrix. 

Apart from the local sparse matrix, each processor might hold additional local data assigned to it. 
This data can be another sparse matrix, a dense matrix, or a vector of elements.
The local data can be abstracted as DUs.

% \begin{tikzpicture}[>=latex]
%   \matrix (A) [matrix of math nodes,left delimiter={[},right delimiter={]}] at (0,0)
%   {
%     a_{11} & 0 & 0 & a_{14} \\
%     0 & a_{22} & 0 & 0 \\
%     0 & 0 & a_{33} & 0 \\
%     0 & 0 & 0 & a_{44} \\
%   };
%   \matrix (B) [matrix of math nodes,left delimiter={[},right delimiter={]}, below=0.5cm of A] at (0,0)
%   {
%     b_{11} & b_{12} & b_{13} & b_{14} \\
%     b_{21} & b_{22} & b_{23} & b_{24} \\
%   };
%   \matrix (C) [matrix of math nodes,left delimiter={[},right delimiter={]}, right=2cm of A] at (0,0)
%   {
%     c_{11} & c_{12} & c_{13} & c_{14} \\
%     c_{21} & c_{22} & c_{23} & c_{24} \\
%     c_{31} & c_{32} & c_{33} & c_{34} \\
%     c_{41} & c_{42} & c_{43} & c_{44} \\
%   };
%   \draw[->] (A) -- (C) node[midway,above] {$A$};
%   \draw[->] (B) -- (A) node[midway,right] {$B^T$};
%   \draw[->] (A) -- (C) node[midway,below] {$C$};
% \end{tikzpicture}

\begin{figure}[ht!]
        \centering
        \includegraphics[width=0.4\columnwidth]{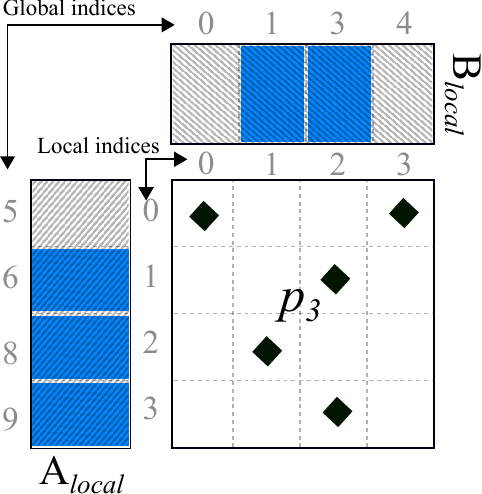}
        \caption{Local SDDMM view of processor $p_3$ (cf. \figurename~\ref{fig:localview}) after applying \fwname localization. The global and local indices are mapped to each other with \gmap and \lmap.}
        \label{fig:localmatrix}
        \vspace{-2ex}
\end{figure}

\subsection{Sparse communication model}
\label{sec:sparseCommModel}

Sparse communication can be very irregular, and is best defined with a point-to-point (P2P) communication graph $commG=(\mathcal{P}, \mathcal{M})$.
While $\mathcal{P}$ is the set of all processors as defined previously, $\mathcal{M}$ is the set of all messages between these processors.
A message $\msgab \in \mathcal{M}$ exists if there is at least one DU that either owned by $p_\alpha$ and required by $p_\beta$, or partially computed by $p_\alpha$ and owned by $p_\beta$.
%Each edge is associated with weight $s(\cdot)$ that resembles the size of the message between two processors.

During \precomm, the messages typically constitute DUs that are owned by one processor and required by others (broadcast).
During this phase, a DU can appear in several outgoing messages, whereas incoming messages contain unique DUs.
This is because each DU is owned by a single processor, and it is not possible to receive it from multiple processors.

During \postcomm, the messages are typically partial results of DUs that will be sent to their respective owners (reduce).
During this phase, a DU can appear in several incoming messages, whereas outgoing messages contain unique DUs.

The distinction whether incoming/outgoing messages contain unique, or otherwise non-unique, DUs is important for the discussion on how to we enable true zero-copy sparse communication in \fwname.

\begin{figure}[h]
\centering
\includegraphics[width=0.8\columnwidth]{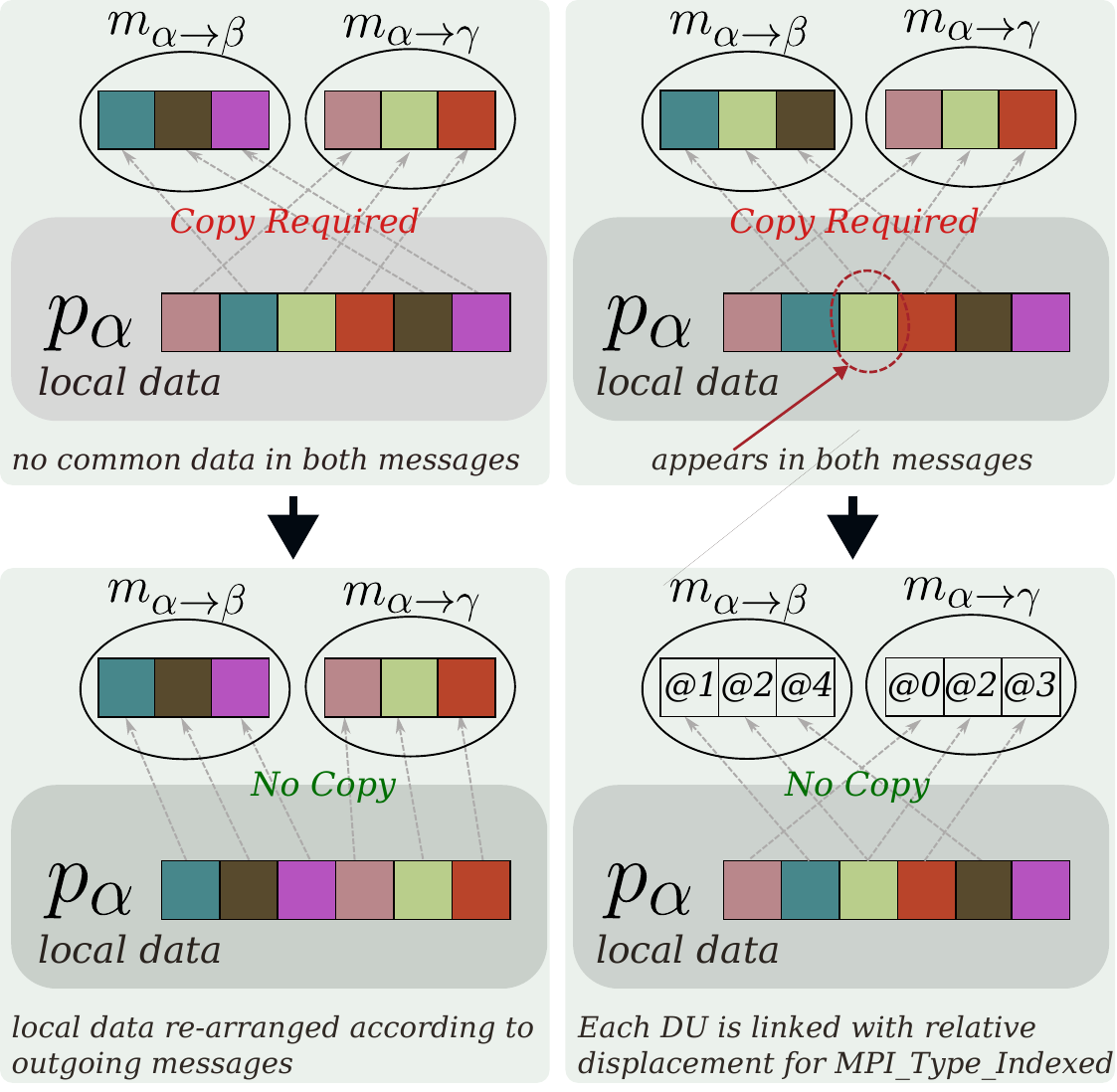}
\label{fig:dataMsg}
\caption{Zero-copy techniques based on DU uniqueness within incoming/outgoing message.}
\vspace{-2ex}
\end{figure}

% If a UoD to be used or updated by different processors, then it will be communicated while executing the sparse kernel.
% There are two types of communication a UoD could be involved in at any given time, either it will be sent to other processors that require it (broadcast), or it requires partial values from other processors (reduce).

% A message $\msgab$ exists if there is at least one UoD owned by 

% We make a data-aware categorization on the incoming or outgoing edges that helps us distinguish and improve the communication in later sections. Data-unique (DU) in/out messages when the UoD associated with a global ID appears only once in messages to/from $outSet(p_\alpha)$/$inSet(p_\alpha)$.}
%     \item data-dup in/out messages: when the UoD associated with a global ID appears multiple times in one or more messages to/from $outSet(p_\alpha)$/$inSet(p_\alpha)$. This typically happens in broadcast or reduce communication.
% \end{itemize}  

\sloppy{The messages are exchanged using either point-to-point communication (\texttt{MPI\_Send/MPI\_Recv} or their non-blocking equivalents).}
We propose to exchange these messages without relying on send/receive buffers.
For the sake of completeness, we propose three versions of handling buffers in sparse communication where the first method assumes the use of send/receive buffers.
%different alternatives for handling buffers in these MPI primitives. 

\subsubsection{Method1: \textbf{Sp}arse \textbf{C}ommunication with \textbf{B}oth \textbf{B}uffers (\textbf{\mbox{SpC-BB}})}

The straight forward way to build the messages in $commG$, given that global IDs to be in the message are known, is to go over each global ID and copy its associated DU to a buffer.
Similarly, at the receiver's end, copy the received DU one by one to their correct location in memory according to their global ID.
This approach adds extra costs when compared to the sparsity-agnostic bulk communication approach, which are the cost of extra memory resembled in the send and receive buffers, as well as the cost of data movement while copying to/from the local data.
This method works regardless of the uniqueness of DUs in outgoing/incoming messages.

\subsubsection{Method2: \textbf{Sp}arse \textbf{C}ommunication with \textbf{S}end/\textbf{R}ecv \textbf{B}uffer Only (\textbf{\mbox{SpC-SB}/\mbox{SpC-RB}})}

Our first improvement over SpC-BB it to get rid of \emph{either} the send \emph{or} the receive buffers by re-arranging how the local data is stored to align with the order of the sent/received data. 
In other words, assume processor $p_\beta$ sends/receives from $n$ processors, $1\!\le \! n \!\le \!Y$, in $P_{x,:,z}$ in the following sequence $\langle p_{i_1}, p_{i_2},\dots, p_{i_n}\rangle$ and the set of DUs received by $p_{i_n}$ is denoted by $\mathcal{D}_{i_n}$.
%
%We also denote the set of owned rows by $\mathcal{R}_{\textit{owned}}$.
Then, the local data at the processor is stored as \[\left[\mathcal{D}_{i_1}, \mathcal{D}_{i_2}, \dots, \mathcal{D}_{i_n} \right]\]
and the messages are directly sent/received at the position starting from the first index of $\mathcal{D}_{i_1}$.

% \begin{equation*}
%     \left[\mathcal{R}_{\textit{owned}}, \mathcal{R}_{i_1}, \mathcal{R}_{i_2}, \cdots, \mathcal{R}_{i_n} \right]
% \end{equation*}

With this method, it is only possible to remove the buffers for either the outgoing or incoming messages, whichever has unique DUs.
This is because, if the messages have non-unique DUs the relation between the duplicates of the DUs in several messages and their unique location in local memory becomes many-to-one rather than one-to-one, hence making the local data alignment impossible.
% the send and receive buffers if the messages to $outSet$ and from $outSet$ are data-unique.
% For data-dup messages, it is not possible to remove the need for the send/receive buffer by simply aligning local data storage with sent/received messages data.

\subsubsection{Method3: \textbf{Sp}arse \textbf{C}ommunication with \textbf{N}o \textbf{B}uffers (\textbf{\mbox{SpC-NB}})}

\sloppy{For messages with non-unique DUs, we utilize \texttt{MPI\_Type\_Indexed} to address the multiple copies of the same DU without the need to explicitly moving them to a buffer.
\texttt{MPI\_Type\_Indexed} creates a new datatype by aligning blocks of pre-defined datatype.
The \texttt{MPI\_Type\_Indexed} type is initialized with the starting address of a contiguous memory chunk, and each block is defined with length and displacement from the starting address.
It is possible to create two (or more) copies/projections of the same local DU in two messages $M_{\alpha \rightarrow \beta}$ and $M_{\alpha \rightarrow \gamma}$ (or more) by simply adding the DU as a block in the \texttt{MPI\_Type\_Indexed} object of each message with the same displacement.}

\sloppy{We consider a DU to be the minimum block in  \texttt{MPI\_Type\_Indexed}.
If multiple DUs appear consecutively in memory, we merge them into a single block in order to minimize the data required to build the \texttt{MPI\_Type\_Indexed} object.

% \sloppy{SpC-NB utilizes \texttt{MPI\_Type\_Indexed} if both $outSet$ and $inSet$ messages are data-dup.
% If, on the other hand, either $outSet$ or $inSet$ messages are data-unique, then it utilizes the alignment method described for SpC-SB.}

% While the actual communication in the communication graph is performed during the \precomm phase, building the communication graph itself is performed once during the \init phase.
% The goal of this separation is to reduce the overhead during the \precomm stage to merely data movement between processors.
% On the other hand, all the other constant information necessary for performing seamless communication, such as in/out sets, message sizes, what to send and where to find the data, receive size from each processor and where to store the received data, and others, are prepared beforehand.

\fwname takes the communication graph from the user in the \init phase. 
Based on this graph, the \fwname framework internally handles all the steps necessary to perform efficient communication during the \precomm phase, including the zero-copy implementations.
\fwname returns a communication object per communication graph in the setup phase.
Assuming the \precomm communication object is called \texttt{precomm}, then during the \precomm phase the user simply calls \texttt{precomm.communicate()} to perform the efficient sparse communication.

% \subsection{Satisfying communication requirement with bulk (dense) communication}
% The easiest way to perform the communication in SDDMM is to let each processor send its owned part of the dense matrix to the other processors in the same row/column of the process grid.

% State of the art methods that compute SDDMM on distributed-memory fashion rely on communicating rows of $A$ and $B$ in bulk using dense communication methods such as broadcast. 
% However, this poses a scalability challenge in terms of communication volume as well as memory footprint, especially for large $\M$, $\N$ and $\K$ values.

% During the pre-communication phase, 

% \subsection{Satisfying communication requirement with sparse communication}

\section{Building 3D SDDMM and SpMM Algorithms with \fwname}
We begin by showing in detail the steps to build 3D SDDMM algorithm with \fwname, and later we reuse some of these steps for 3D SpMM. 
%
% We assume the sparse matrix $\Smat$ is distributed into the 3D $X\times Y \times Z$ grid according to \dist. 
% Processor $p_{xyz}$ holds $\Smat_{xyz}$ with $nnz(\Smat_{xyz})$ nonzero elements, $nrows(\Smat_{xyz})$ local rows and $ncols(\Smat_{xyz})$ local columns. The local rows and column counts do not include any empty row/column resulting from the distribution.
% Then, each local SDDMM computation would require a local  matrix $\Amat^z_{x}$ of size $nrows(\Smat^z_{x,z})\times \frac{\K}{Z}$ and a local matrix $\Bmat^z_{y}$ of size $ncols(\Smat^z_{x,y}) \times \frac{\K}{Z}$.
%
Using \dist, we can formulate the 3D parallel SDDMM algorithm as follows: the $\Amat$ and $\Bmat$ matrices are partitioned row-wise into $X$ and $Y$ parts conformably with the rows and columns of $\Smat$, respectively.
Then, $\Amat$ and $\Bmat$ are partitioned column-wise to $Z$ parts $\Amat^1,\cdots, \Amat^z, \cdots,\Amat^Z$.
Here, a part $\Amat^z_{x}$ is used by processors in $P_{x,:,z}$.
Similarly, a part $\Bmat^z_{y}$ is used by processors in $P_{:,y,z}$
We assume each processor in $P_{x,:,z}$ owns almost equal number of rows of $\Amat^z_x$.
This naturally means that $\Amat^z_x$ is further divided into $Y$ parts. 
We abuse the 2D notation of matrix blocks to refer to 1D internal parts of a dense 1D matrix block.
For instance, $\Amat^z_{x,y}$ refers to the 1D block $\Amat^z_{xX+y}$.
The processor that owns a row $\abold^z_i$ is retrieved with the $owner(\abold^z_i)$.

We start by defining the requirements of the \comp phase, and then move to explain how to fulfill these requirements in the \init, \precomm, and \postcomm phases.

\subsection{The Compute phase}

In 3D SDDMM, the responsibility of a processor $P_{x,y,z}$ is to compute partial results of inner products for all nonzeros in $\Cmat_{x,y}$, and then be responsible for holding the final results of nonzeros in $\Cmat^z_{x,y}$.
Here, the two vectors in the inner product $\langle \abold_i, \bbold_j \rangle$ that take place for each nonzero element in $\Smat_{x,y}$ are of size $\frac{\K}{Z}$.
This clearly resembles the shift from 2D to 3D algorithms as dividing the fine-grain tasks of computing the inner product of two dense vectors of size $\K$ into $Z$ sub-tasks each of size $\frac{\K}{Z}$.

Initially, $P_{x,y,z}$ owns $\Smat^z_{x,y}$ only.
Therefore, it requires \mbox{$\Smat_{x,y} \setminus \Smat^z_{x,y}$} as well as all (sub)rows of $\Amat$ and $\Bmat$ required for the computation of $\Cmat_{x,y}$.
With the assumption that the values and sparsity patter of $\Smat_{x,y}$ are constant during multiple iterations of computing SDDMM, the gathering of $\Smat_{x,y}$ can be done at the \init phase.
On the other hand, the values in $\Amat$ and $\Bmat$ will change at each iteration, therefore they should be communicated at the \precomm phase.

The local compute phase at each processor is stand-alone by the design of the framework.
A processor $P_{x,y,z}$ has a localized version of $\Smat_{x,y}$ as well as all required $\Amat$ and $\Bmat$ rows by the end of the \precomm phase.
\figurename~\ref{fig:localmatrix} shows an example of the state of local SDDMM.
The \comp phase is performed following~\eqref{eq:sddmm}, either on CPU, GPU or any accelerator for sparse computations using state-of-the-art sequential or shared-memory-parallel codes.

\subsection{The PreComm phase}
\label{sec:sddmmprecomm}

In this phase, the dense $\Amat$- and $\Bmat$-matrix rows are communicated.
Computing partial $\Cmat_{x,y}$ by $P_{x,y,z}$ requires dense rows from $\Amat^z_x$ and $\Bmat^z_y$.
Since $\Amat^z_x$ is required by all processors in $P_{x,:,z}$, we assume that $P_{x,y,z}$ owns an equal part of $\Amat^z_x$. Similar discussion hold for $\Bmat^z_y$.

Since the core goal of \fwname is to perform sparse communication, we follow the $\lambda$-based communication discussed in Section~\ref{sec:SDDMMComm}.
A message $\msgab$ from processor $p_\alpha$ to processor $p_\beta$, where $p_\alpha, p_\beta \in P_{x,:,z}$ is formed as
\begin{equation}
\label{eq:commGi}
    \msgab = \{\abold_i \mid p_\alpha,p_\beta \in \Lambda_i \land owner(\abold_i) = p_\alpha\}.
\end{equation}
Similarly, a message $m_{\gamma \rightarrow \delta}$ from processor $p_\gamma$ to processor $p_\delta$, where $p_\gamma, p_\delta \in P_{:,y,z}$ is formed as
\begin{equation}
\label{eq:commGj}
    m_{\gamma \rightarrow \delta} = \{\bbold_j \mid p_\gamma,p_\delta \in \Lambda_j \land owner(\bbold_j) = p_\gamma\}.
\end{equation}
%
%
% \sloppy{The sparse communication messages built using~\eqref{eq:msg} are exchanged using either point-to-point communication (\texttt{MPI\_Send/MPI\_Recv}), or using neighborhood collectives, specifically \texttt{MPI\_Neighbor\_alltoallv}.}
%For both techniques, the data in $M_{\alpha, \beta}$ should be stored 
%

%preliminary step prior to starting the main communicate-compute loop.
% The goal of this step is to setup all the information necessary for performing seamless communication, from determining the message sizes, what to send and where to find the data, receive size from each processor and where to store the received data, and others.

% \subsection{The PostComm phase}

% In this phase, each processor receives partial results for the nonzeros it owns from other $Z\!-\!1$ processors.
% While the communication in this phase can be formulated as sparse communication similar to the \precomm phase, we prefer to perform it with a Reduce-Scatter call since the cost of communication in this phase is negligible compared to \precomm.

\subsection{The PostComm phase}

Gathering the partial results of $\Cmat_{x,y}$ requires that processor $P_{x,y,z}$ receives all partial results of the nonzero elements it owns from the other $Z-1$ processors in $P_{x,y,:}$.
This amounts to receiving $(Z-1)\times nnz(\Smat_{x,y,z})$ words of data.
With highly sparse matrices, the cost of this phase is very low compared to \precomm. 
We perform this operation with a Reduce-Scatter rather that converting it to sparse communication primitives.

\subsection{The Setup phase}

All the configurations required for the \precomm, \comp, and \postcomm phases are performed in this phase.
These configurations are performed once and used multiple times.
The first is gathering $\Smat_{x,y}$ at each processor in $P_{x,y,:}$. 
This is done with an All-Gather operation on $\Smat^z_{x,y}$ by all processors in $P_{x,y,:}$.

%\subsubsection{$\lambda$-aware distribution of $\Amat$ and $\Bmat$}
The second configuration is distributing the dense $\Amat$ and $\Bmat$ among processors.
In Section~\ref{sec:sddmmprecomm} we stated our assumption that the dense rows are distributed equally (owned) to the processors that use them. 
However, without careful attention, a random equal distribution will invalidate the $\lambda$-based communication discussed in Section~\ref{sec:SDDMMComm}.

%\subsection{$\lambda$-aware data distribution}
%\label{sec:lambdaawaredist}

The $\lambda$-based communication of a dense row $\abold_i$ is accurate only if the processor that owns $\abold_i$ is part of $\Lambda$.
Otherwise, a dense row $\abold_i$ that is assigned to a processor outside of $\Lambda_i$ will incur an extra unnecessary communication and storage of size $K$ words.

\begin{algorithm}
\caption{Parallel $\lambda$-aware random distribution on $p_\alpha$}
\label{alg:commResp}
\flushleft \hspace*{\algorithmicindent} \textbf{Input} $nrows(\Smat)$ or $ncols(\Smat)$ as $gsize$, $nrows(\Smat_\alpha)$ or $nrows(\Smat_\alpha)$ as $dsize$, $P$, \gmap\\
\flushleft \hspace*{\algorithmicindent} \textbf{Output} owner array $owner$
    \begin{algorithmic}[1]
        \Require{}
        \State{$myrows$ $\gets$ the $myrank$ chunk of $dsize/P$}
        \State{candidates $\gets$ array of lists of size $|myrows|$}
        \State{sendInfo $\gets$ array of lists of size $P$}
        \For{$i$ in range $(1, ldsize)$}
            \State{$gi \gets$ globalMap[$i$]}
            \State{$p \gets$ processor responsible for assigning $gi$}
            \If{$p = myrank$}
            \State{candidate[$gi$] $\gets$ candidates[$gi$] $\cup$ $p$}
            \Else
            \State{sendInfo[$p$] $\gets$ sendInfo[$p$] $\cup$ $gi$}
            \EndIf
        \EndFor
        \State{recvInfo $\gets$ Exchange sendInfo}
        \For{$p$ in range $(1, P)$}
            \For{row ID $rid$ in recvInfo[$p$]}
                \State{candidates[$rid$] $\gets$ candidates[$rid$] $\cup$ $p$}
            \EndFor
        \EndFor
        \State{$myowner \gets$ empty array of size $|myrows|$}
        \For{row ID $rid$ in $myrows$}
            \State{$myowners[rid] \gets$ pick a random $p \in$ candidates[$rid$]}
        \EndFor
        \State{All-Gather($myOwner$, $owner$)}
        
    \end{algorithmic}
\end{algorithm}

We propose Algorithm~\ref{alg:commResp} to efficiently distribute the dense rows to processors in their respective $\Lambda$ sets in parallel.
The algorithm starts by dividing the work to be done by each processor.
Each processor is responsible for finding an owner of a set of rows, and these rows can be assigned randomly in a load balanced manner (\emph{lines 1-2}).
Then, each processor loops over the rows it uses and send their global index to the processor responsible for their assignment (\emph{lines 4-13}).
After each processor receives the list of candidates to each row, it loops over the candidates and picks one processor at random to assign it as the owner of the corresponding row (\emph{lines 14-22}).
The candidates that each processor receives for each row index are the processors that actually use that index in their local computation, i.e., they have at least one nonzero element with such an index. 
For this reason, the processor that is picked at random here will certainly be one of the processors involved in the communication of the target row, thus not incurring any additional unnecessary communication.
Finally, an All-Gather operation is performed in order to gather the ownership information from all processors (\emph{line 23}).

The third configuration that will be used multiple times during the parallel SDDMM is building the communication graph of the \precomm phase.
Based on this graph, the \fwname framework internally handles all the steps necessary to perform efficient communication during the \precomm phase, including the zero-copy implementations.

\subsection{Building SpMM with \fwname}
The process to building SpMM is similar to that of SDDMM.
The distribution of the sparse matrix is the same as that of SDDMM.
In the \comp phase, a processor $P_{x,y,z}$ is responsible for computing partial results of dense rows in $A^z_{x,y}$.
For that, $P_{x,y,z}$ also needs the whole $\Smat_{x,y}$, which is gathered during the \init phase.
The SpMM requires \mbox{$\Bmat$-matrix} rows in the \precomm phase. 
The communication graph is built with~\eqref{eq:commGj}.
After the \comp phase, $P_{x,y,z}$ sends partial results for all the dense rows that it does not own to their respective owners during the \postcomm phase.
The communication graph in this phase is constructed with~\eqref{eq:commGi} but replacing $owner(\abold_i)=p_\alpha$ by $owner(\abold_i)=p_\beta$.
This is because the owner is the receiving party not the sending party as this is a reduce operation.
The communication cost of the \precomm phase of SDDMM can be thought of as distributed among both \precomm and \postcomm phases in SpMM. 
In other words, both \precomm and \postcomm phases in SpMM are of equal importance, unlike SDDMM.

\section{Experimental Evaluation}
\label{sec:exp}

Our evaluation goal is to empirically assess the theoretical aspects of \fwname in terms of reducing communication and memory footprint, and how they reflect on actual runtime and scalability, compared to sparsity-agnostic 3D algorithms.

In line with the presentation of the paper, we consider parallel SDDMM and SpMM in our experiments.
We use real-world sparse matrices with at least 100M nonzero elements, and we vary three variables, $K$, $Z$, and $P$ to study their effect on both \fwname and the sparsity-agnostic 3D algorithms.
Unless stated otherwise, all runtime results reported in this section are result of averaging five different runs per experiment.

For the sparsity-agnostic 3D algorithm (\S~\ref{sec:comm}), we provide our own implementation (referred to as \densename hereafter), and also compare against the state-of-the-art framework Half-and-Half (\hnhname) by Bharadwaj et al.~\cite{hnh2022} that provides the same algorithm under the name "2.5D sparse replicating" (referred to as \hnhname hereafter).
This existing framework is compared against PETSc~\cite{petsc-user-ref} and shown to outperform it significantly. 
For this reason, we refrain from comparing against PETSc. 
We use the abbreviations in Section~\ref{sec:sparseCommModel}: SpC-BB, SpC-R/SB, and SpC-NB to refer to the specific implementation in \fwname, and we use the framework's name when comparing metrics irrelevant to the implementation such as communication volume.

\subsection{Dataset and Experimental Setting}

\begin{table}[!h]
\setlength{\tabcolsep}{8pt}
  \centering
  \caption{Sparse matrices used in our experiments}
  \label{tab:datastats}
  \setlength\tabcolsep{6pt}
   %\resizebox{\columnwidth}{!}{
   \begin{tabular}{lrrrrr}
     \toprule
    Matrix & \#rows/cols & \#nonzeros &  Density 
    %$\left(\frac{nnz^2}{nrows^2}\right)$
    \\
    \midrule
%GAP-road	&	23,947,347	&	23,947,347	&	57,708,624	&	9	&	9	&	1.01$\times 10^{-07}$ \\

arabic-2005	&	22,744,080	&	639,999,458	& 1.24$\times 10^{-06}$	\\
delaunay\_n24	&	16,777,216	&	100,663,202	& 3.58$\times 10^{-07}$	\\
europe\_osm	&	50,912,018	&	108,109,320	&	4.17$\times 10^{-08}$\\	
GAP-kron	&	134,217,726	&	4,223,264,644 & 2.34$\times 10^{-07}$ \\	
GAP-road	&	23,947,347 &	57,708,624	&1.01$\times 10^{-07}$	\\
GAP-web	&	50,636,151	&	1,930,292,948	&	7.53$\times 10^{-07}$	\\
kmer\_A2a	&	170,728,175	&	360,585,172	&	1.24$\times 10^{-08}$	\\
twitter7	&	41,652,230 &	1,468,365,182	& 8.46$\times 10^{-07}$\\
uk-2002	&	18,520,486	&	298,113,762	&	8.69$\times 10^{-07}$\\	
webbase-2001	&	118,142,155	&	1,019,903,190	&	7.31$\times 10^{-08}$\\

    \bottomrule
\end{tabular}
%}
\end{table}

We evaluate with ten real-world sparse matrices that represent graphs, obtained from the The SuiteSparse Matrix Collection\footnote{https://sparse.tamu.edu/}~\cite{sparseSuiteMatrixCollection}.
The properties of the sparse matrices are detailed in \tablename~\ref{tab:datastats}.
All our matrices have between 100 million and 4.2 billion nonzero elements.

We implemented \fwname as well as \densename using C++ and used MPI for inter-process communication.
All our experiments are taken on CSCS Piz Daint HPC system based in Switzerland.
We use the CPU partition of the Cray XC40/XC50 system, which is equipped with 1813 dual-socket Intel Xeon E5-2695 processors clocked at 2.10GHz and has 64GiB of DD3 RAM memory.
The nodes are connected with a Cray Aries network that uses a dragonfly network topology.
All the codes are compiled with a Cray-clang compiler and a system-provided Cray-MPICH on SUSE Linux Enterprise Server 15-SP2 operating system.

\subsection{High-level total runtime comparison with sparsity-agnostic algorithms}

We begin by comparing \fwname against \densename and \hnhname in terms of total runtime of five iterations of SDDMM followed by SpMM.
The reason for this comparison is that \hnhname is designed to perform fused SDDMM and SpMM operations, and has restrictions when it comes to choosing $Z$ and $K$ values. 
Also, the $X$ and $Y$ dimensions of the 3D mesh should be equal in \hnhname.
\fwname does not have such restrictions, and works with any $X$, $Y$, $Z$, and $K$ values as long as the trivial restrictions hold: $K/Z\!\ge\!1$ and $X\times\!Y\!\times\! Z \!= \!P$.
For this reason, we set $Z$ and $K$ parameters to values that work for \hnhname, which are respectively 4 and 60.
We later vary these values in experiments that do not include \hnhname.

\begin{figure}[h]
    \centering
    \resizebox{\linewidth}{!}{
    \includegraphics{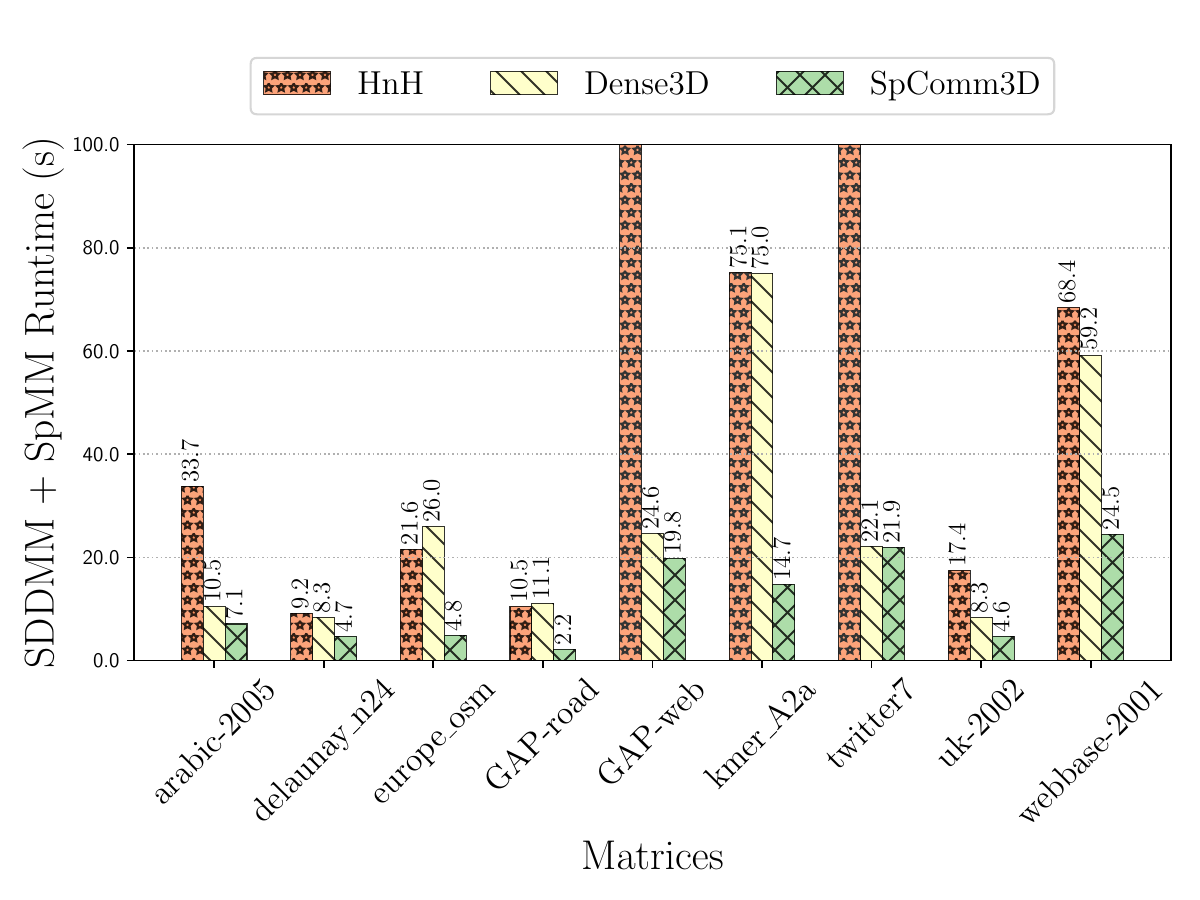}}
    \caption{Comparing \fwname, \densename, and HnH in terms of five SDDMM followed by SpMM operations runtime.}
    \label{fig:compareHnH}
    
\end{figure}

\figurename~\ref{fig:compareHnH} compares \fwname, \densename, and HnH in terms of runtime of five SDDMM followed by SpMM operations on 900 processors.
HnH as is without any modifications by selecting the proper 2.5D algorithm ``2.5D sparse replicating''.
As the figure shows, \fwname  significantly outperforms both \densename and \hnhname.
Although theoretically the same algorithm, \hnhname and \densename show different runtimes on different matrices.
While most instances show that both methods perform similarly, or slightly in favor of one of them, three instances show significant difference in favor of \densename.
%We believe that the reason for this is that \hnhname uses a different sparse matrix distribution as well as different dense row/column distribution.
This behavior could be explained by the fact that  \hnhname uses multiple blocking send/recieve calls (\texttt{MPI\_Sendrecv}) per processor to realize the All-Gather operation required for the first phase of communication, whereas \densename uses non-blocking broadcasts (\texttt{MPI\_Ibcast}) to realize the same communication.

\subsection{Strong Scaling}

\begin{figure*}[!ht]
    \centering
    \includegraphics[width=\linewidth]{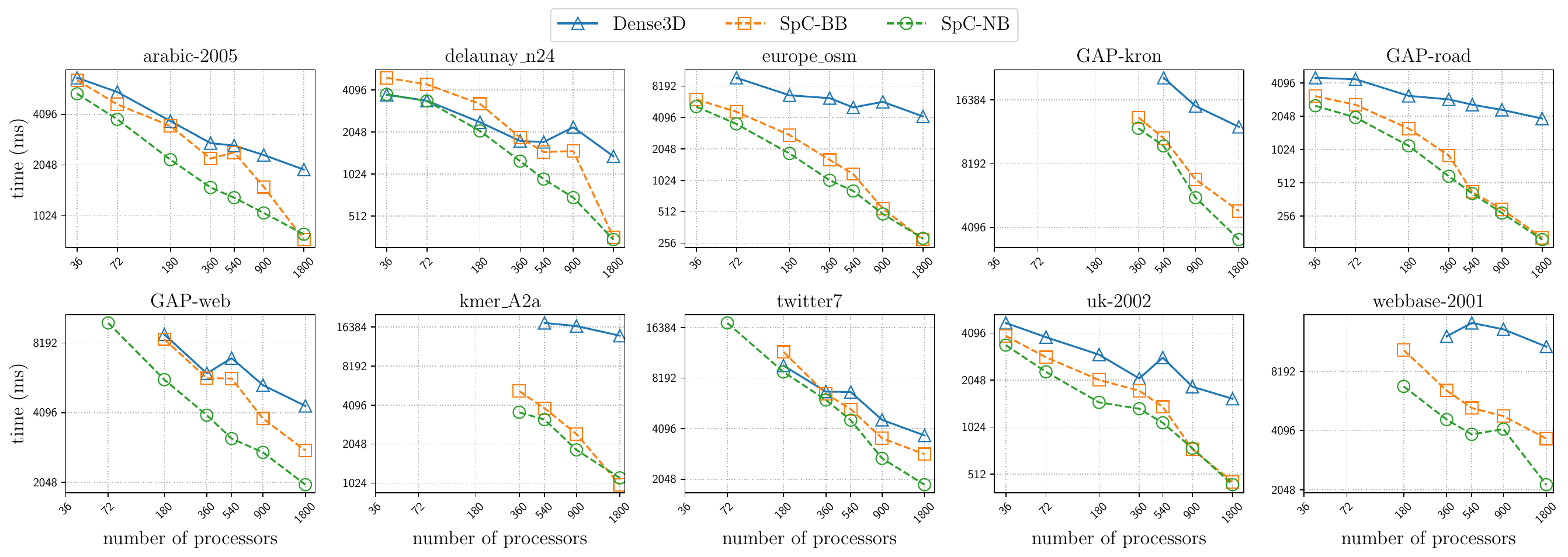}
    \caption{Strong Scaling (log-log) of SDDMM with $K$=120 and $Z$=4. A missing value means infeasible run (out-of-memory).}
    \label{fig:ss}
    \vspace{-2ex}
\end{figure*}

In order to study the scaling of SDDMM with \fwname compared to \densename, we run SDDMM on 36, 72, 180, 360, 540, 900, and 1800 processors.
Within \fwname, we compare two implementations of sparsity-aware communication: SpC-BB and SpC-NB.
We fixed $K$ value to 120 and $Z$ value to $4$.
\figurename~\ref{fig:ss} shows the strong scaling results for all matrices in our dataset.
In the figure, there are some missing values that we were unable to obtain due to high memory demand (out-of-memory errors).
Most of such missing values are from the \densename method, and/or with experiments taken on small number of processors ($P$=180 or below).

As seen in the figure, sparsity-aware communication is superior to \densename in terms of runtime and memory scalability.
There is a significant gap between \densename and the other sparsity-aware methods in terms of runtime, especially as $P$ becomes higher.
When comparing sparsity-aware methods against each other, it is clear that SpC-BB is inferior to SpC-NB in all cases when $P < 900$.
When $P$ grows larger than 900, SpC-BB performs similar to SpC-NB in six out of ten matrices.

%In terms of memory feasibility, were were able to complete runs for all methods on all processor counts only for ``arabic-2005'', ``delaunay\_n24'', and ``uk-2002''.
%For others, there were at least one memory-infeasible run.
%For ``europe\_osm'', sparsity-aware methods are run on all processors, whereas \densename was infeasible on $P$=32.
%For ``GAP-twitter'' and ``GAP-web'' runs started to be feasible from $P$=72 processors with SpC-RB abd SpC-NB, whereas for \densename and SpC-BB the runs started to be feasible starting from $P$=180.
%For the rest of the matrices, runs started to be feasible at $P$=360 or later.
%The figures show that sparsity-aware methods show much better memory feasibility compared to \densename.
%Furthermore, buffer-avoiding methods (SpC-RB, SpC-NB) show better feasibility in two out of ten cases.

\subsection{Harnessing Sparsity: evaluation of memory and communication}

\begin{figure}[!h]
    \centering
    \begin{subfigure}[b]{\columnwidth}
    \centering
    \includegraphics[width=\columnwidth]{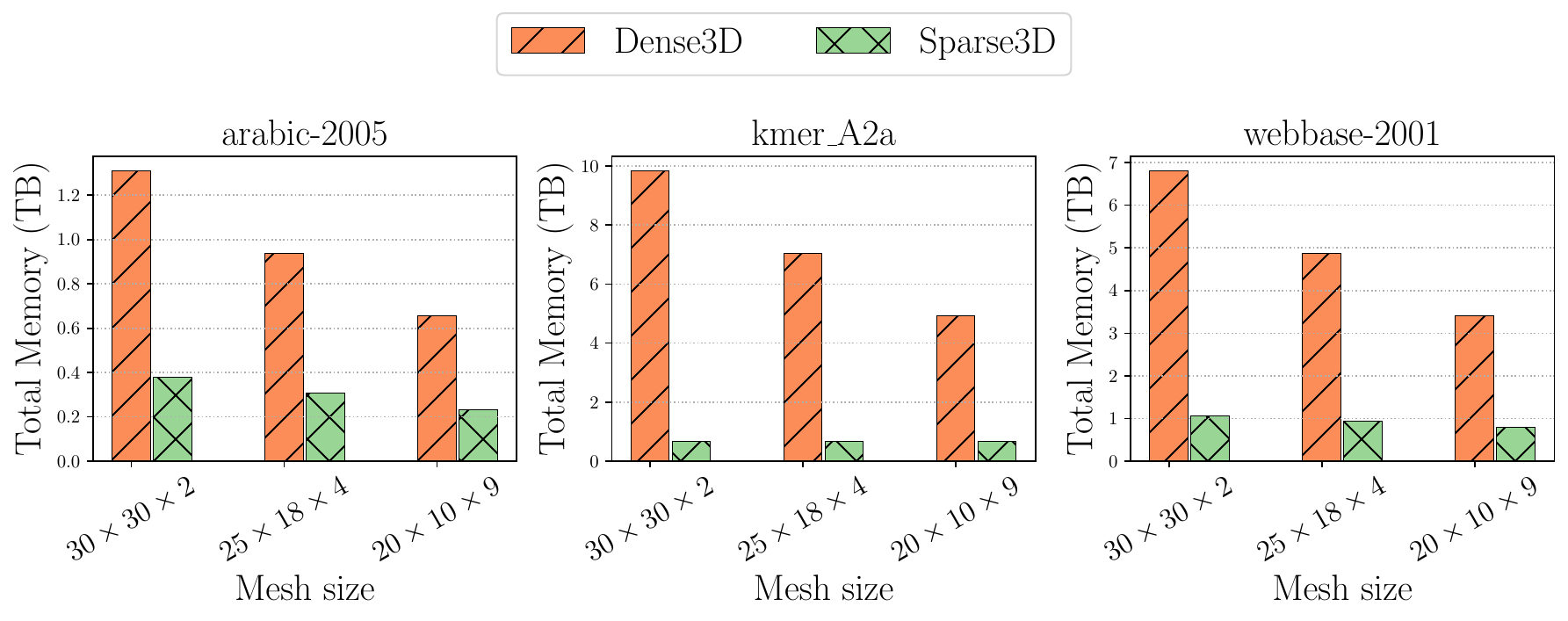}
    \end{subfigure}

    \begin{subfigure}[b]{\columnwidth}
    \centering
    \includegraphics[width=\columnwidth]{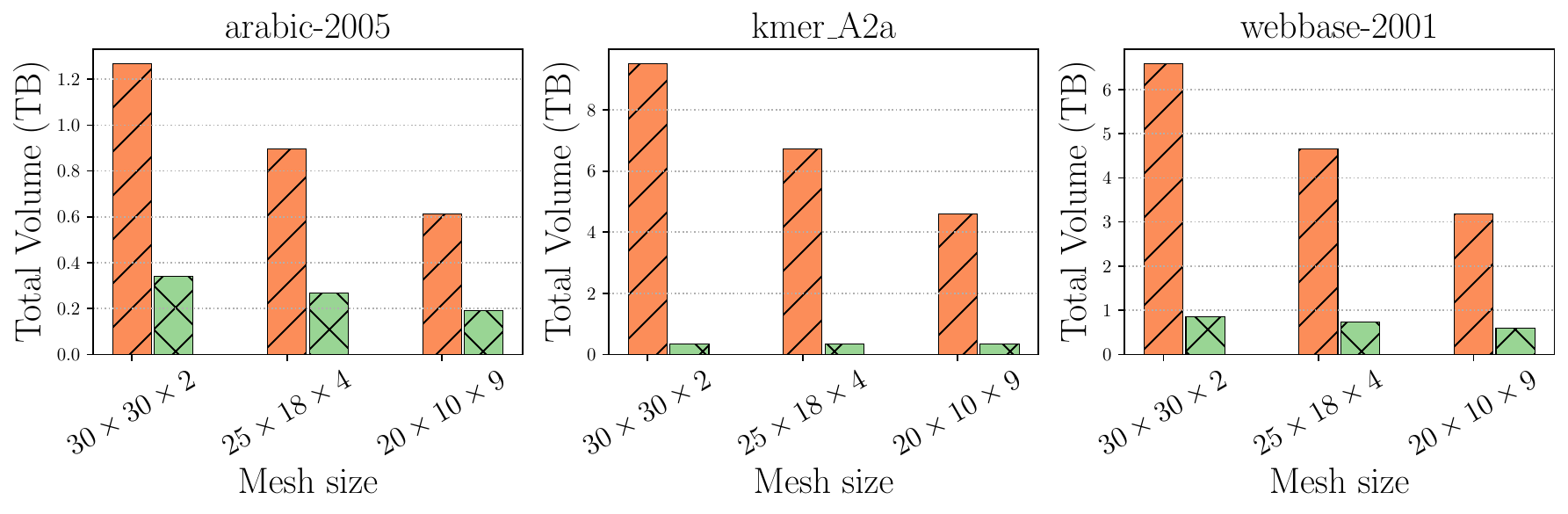}
    \end{subfigure}
     
     \begin{subfigure}[b]{\columnwidth}
    \centering
    \includegraphics[width=\columnwidth]{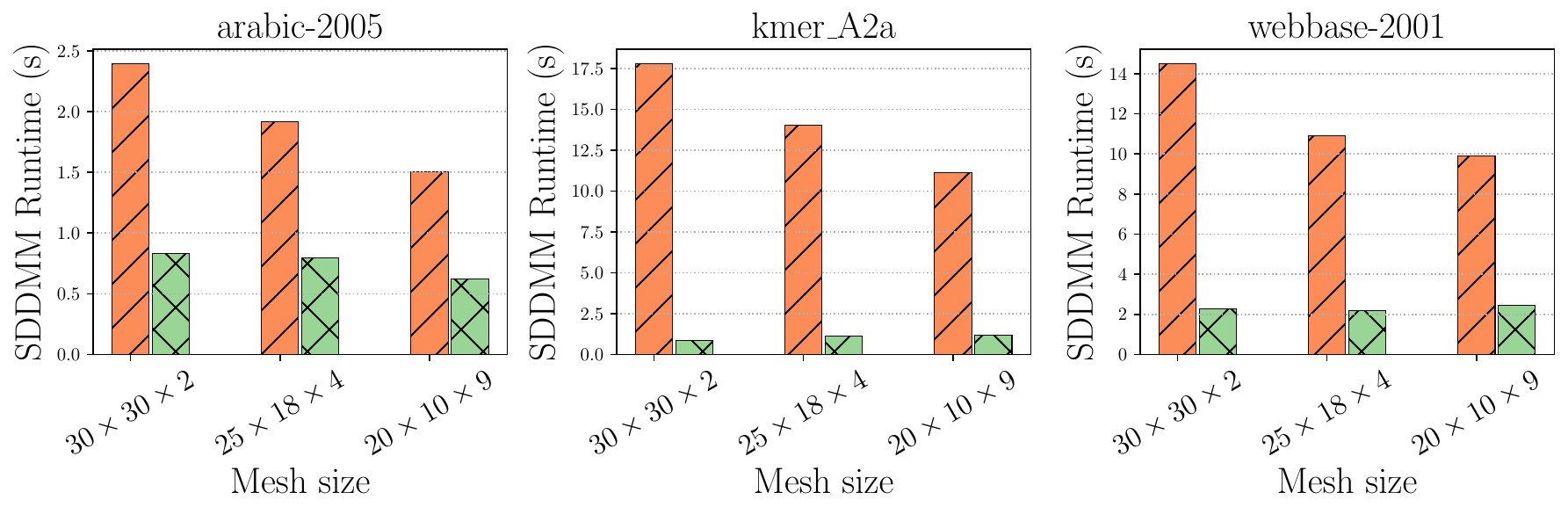}
    \end{subfigure}
    
    \caption{Memory, volume, and SDDMM runtimes of \densename vs. \fwname on three matrices on 1800 processors (with $K=120$).}
    \label{fig:memGB}
    \vspace{-2ex}
\end{figure}

As demonstrated in this paper, using sparsity-aware communication in \fwname not only reduces the actual volume of communication between processors, but also enables reducing the overall memory footprint by avoiding the storage of any unnecessary dense $\Amat$ or $\Bmat$ rows.
We have measured the total memory required for storing dense $\Amat$ and $\Bmat$ on 1800 processors when $K$ value is set to 240.
We used three matrices from our dataset, and varied the $Z$ dimension between 2, 4, and 9, leading to three different 3D mesh configurations. 
\figurename~\ref{fig:memGB} shows that the overall memory consumption of \fwname is significantly lower than that of Dense.
On ``arabic-2005'', the memory is reduced by 2.5x to 3.5x, depending on the value of $Z$.
On ``kmer\_A2a'', the memory is reduced by 5x to 10x, effectively saving 4TB to 10TB of overall memory. 
Similar ratios appear with ``webbase-2001''.
The memory consumption of \densename is reduced with increasing $Z$ value, an expected behavior following the theoretical analysis.
The memory consumption of the sparsity-aware algorithm also decreases in a slower rate with increasing $Z$ compared to \densename.

\subsection{Overall communication volume and runtime improvements}

\begin{table}[!h]
  \centering
    \caption{Comparing \densename vs all methods of \fwname in terms of max receive volume and total SDDMM runtime on 900 processors with different $Z$ and $K$ values.}
    \label{tab:gm}
  \setlength\tabcolsep{3pt}
    \begin{tabular}{rlrrrr}
		\toprule
		 & & \multirow{3}{*}{\thead[c]{\textbf{Max. Recv}\\\textbf{Volume}\\\textbf{($K$-normalized)}}}&\multicolumn{3}{c}{\textbf{SDDMM runtime (ms)}}\\[1ex]
		% & &\multirow{3}{*}{\thead{Max. Recv volume ($K$-normalized)}}&\multicolumn{3}{c}{\textbf{SDDMM runtime (ms)}}\\[1ex]
		\cmidrule(l{2pt}r{2pt}){4-6}
        $Z$ & \textbf{Method} && $K$= 60 & 120 & 240\\

\midrule
% \multirow{2}{*}{$Z$=2}
% &	Dense3D	&	2,394,372	&	3,419	&	6,860	&	8,713	\\
% &	SpC-NB	&	388,115	&	861	&	1,550	&	3,113	\\
% &	Improvement	&	6.2x &	4.0x	&	4.4x	&	2.8x	\\
%  \vspace{1ex}
%  \\[-6pt]

% \multirow{2}{*}{$Z$=4}
% &	Dense3D	&	1,799,824	&	3,356	&	6,616	&	8,460	\\
% &	SpC-NB	&	252,747	&	877	&	1,527	&	3,089	\\
% &	Improvement	&	7.1x	&	3.8x	&	4.3x	&	2.7x	\\
%  \vspace{1ex}
%  \\[-6pt]

% \multirow{2}{*}{$Z$=9}
% &	Dense3D	&	1,219,179	&	2,348	&	5,010	&	7,617	\\
% &	SpC-NB	&	213,929	&	845	&	1,548	&	3,103	\\
% &	Improvement	&	5.7x&	2.8x	&	3.2x	&	2.5x\\

 \multirow{4}{*}{$Z$=2}
&	Dense3D	&	2,129,152	&	3,021	&	6,063	&	7,798	\\
  \cdashline{2-6}
&	SpC-BB	&	\multirow{3}{*}{328,264}&	795	&	1,600	&	3,597	\\
&	SpC-RB	&	&	720	&	1,434	&	2,702	\\
&	SpC-NB	&	&	714	&	1,295	&	2,587	\\
  \cdashline{2-6}
&	Improvement 	&	\textbf{6.5x} &	\textbf{4.2x}	&	\textbf{4.7x}	&	\textbf{3.0x}\\
  \vspace{1ex}
  \\[-10pt]
  \cmidrule{2-6}
  \multirow{4}{*}{$Z$=4}
&	Dense3D	&	1,453,978	&	2,569	&	5,065	&	7,506	\\
  \cdashline{2-6}
&	SpC-BB	&	\multirow{3}{*}{289,876}	&	902	&	1,773	&	3,772	\\
&	SpC-RB	&		&	732	&	1,407	&	2,958	\\
&	SpC-NB	&	&	785	&	1,374	&	2,771	\\
  \cdashline{2-6}
&	Improvement&	\textbf{5.0x	}&	\textbf{3.3x	}&	\textbf{3.7x	}&	\textbf{2.7x} \\
  \vspace{1ex}
  \\[-10pt]
  \cmidrule{2-6}
  \multirow{4}{*}{$Z$=9}
&	Dense3D	&	981,686	&	1,818	&	3,863	&	6,759	\\
\cdashline{2-6}
&	SpC-BB	&	\multirow{3}{*}{250,387}&	902	&	1,820	&	3,723	\\
&	SpC-RB	&		&	811	&	1,626	&	2,718	\\
&	SpC-NB	&	&	750	&	1,395	&	2,872	\\
\cdashline{2-6}
&	Improvement 	&	\textbf{3.92x	}&	\textbf{2.42x	}&	\textbf{2.77x}	&	\textbf{2.35x}\\

 \vspace{1ex}
 \\[-6pt]
    \bottomrule
    \end{tabular}%
\vspace{-2ex}

\end{table}%

In order to assess \fwname more thoroughly, we report our SDDMM runs using \densename, SpC-BB, SpC-RB, and SpC-NB with $Z=\{2,4,9\}$ and $K=\{60,120,240\}$ on 900 and 1800 processors.
\tablename~\ref{tab:gm} summarizes the runs on 900 processors.
A value in the table represents the geometric average of runs along all matrices in the dataset for the respective metric, method, $K$, and $Z$ values.
The metrics we consider are maximum receive volume and SDDMM running time.
Since the different sparse communication methods within \fwname perform the same communication, all of them share the same max receive volume.
We report volume values normalized with respect to $K$.
We also report the improvement of SpC-NB with respect to \densename for each different $Z$ value.

As seen in the table, \fwname improves the maximum receive volume by 4.0x to 6.5x, depending on $Z$.
This reduction directly impacts the actual runtime, with improvements from 2.35x up to 4.7x, depending on $K$ and $Z$.
Compared against each other, SpC-BB, SpC-RB, and SpC-NB show expected behaviour. 
SpC-NB performs best overall, followed by SpC-RB.
The gap between SpC-BB and the other two methods is clear, especially when $K$ gets higher, whereas SpC-NB and SpC-RB are comparable. 
%As also observed previously, the receive volume of \densename drops by 2x with $Z$ increased from 2 to 4, and by 1.5x from 4 to 9.
%On the other hand

\begin{figure}
    \centering
    \includegraphics[width=0.5\columnwidth]{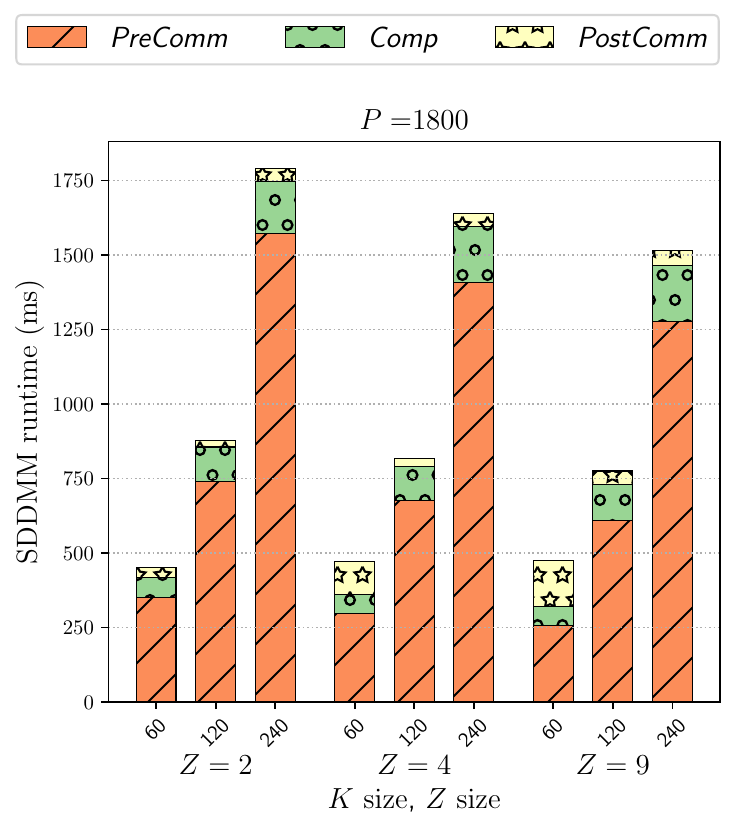}
    \caption{Runtime breakdown of SDDMM with SpC-NB of all matrices on 1800 processors (averaged with geometric mean) categorized according to $K$ and $Z$.}
    \label{fig:gmBar}
    \vspace{-2ex}
\end{figure}

\figurename~\ref{fig:gmBar} shows a breakdown of the SDDMM running times of SpC-NB on 1800 processors with different $K$ and $Z$ values.
The figure shows that \precomm phase dominates the running time.
The \comp phase's share increase as $K$ increases, whereas the share of \postcomm phase increase as $Z$ increases.
 
% \begin{figure*}
%     \centering
%     \includegraphics[width=\linewidth]{plots/sparseNB_bar.pdf}
%     \caption{The effect of changing $Z$ and $K$ on the total runtime of the \textbf{SpC-NB} algorithm on $P=900$ processors. Each plot shows the runtime segments for three different $K$ values: 60, 120 and 240 as well as three different $Z$ values: 2, 4 and 8. Changing $Z$ renders a different mesh size. \todo{add y axis names !!!, reorder the legend }}
%     \label{fig:ss}
% \end{figure*}

\section{Conclusions}
\fwname provides a high-level and flexible environment to run different types of 2D and 3D sparse kernels utilizing sparse communication. 
With extensive experimental evaluations on up to 1800 processors, we showed that \fwname has superior scalability compared to the state-of-the-art sparsity-agnostic SDDMM and SpMM algorithms.
The communication and memory required to run the same algorithm are reduced by an average of 5.0x, leading to saving up to 10TB of bandwidth and RAM on 1800 processors, by effectively utilizing sparse communication.
The overall runtime of SDDMM is also reduced by 5.0x on average.  
The low memory and communication overheads of \fwname compared to the sparsity-agnostic counterparts make it a better candidate for performing large-scale sparse computations especially with very large sparse matrices.

\label{sec:conclusion}
\bibliographystyle{ACM-Reference-Format}
\bibliography{spcomm3d}

\end{document}